\documentclass[conference]{IEEEtran}
\IEEEoverridecommandlockouts

\usepackage{amsmath,amssymb,amsfonts}
\usepackage{graphicx}
\usepackage{algorithm}
\usepackage{algpseudocode}
\usepackage[numbers,square,sort&compress]{natbib}
\usepackage{mathptmx}
\usepackage{times}
\usepackage{tikz}
\usetikzlibrary{arrows.meta,calc}
\usepackage{url}
\usepackage[hidelinks]{hyperref}
\DeclareMathAlphabet{\mathcal}{OMS}{cmsy}{m}{n}

\newtheorem{assumption}{Assumption}
\newtheorem{theorem}{Theorem}
\newtheorem{remark}{Remark}
\newenvironment{proof}[1][Proof]{\noindent\textit{#1:}\quad}{\hfill$\square$\vspace{1ex}}

\begin{document}

\title{PAC-Bayesian Controller Learning for Nonlinear Systems with Unknown Dynamics}
\author{
Yujia Luo, Ye Pu, Jonathan H. Manton, and Jingge Zhu%
\thanks{A shorter version of this work has been submitted
to the 2027 European Control Conference (ECC 2027).}
\thanks{All authors are with The University of Melbourne, Australia.
Emails: \{yujia.luo.1, ye.pu, jmanton, jingge.zhu\}@unimelb.edu.au.}
}

\maketitle

\begin{abstract}
This paper investigates data-driven controller learning for discrete-time nonlinear systems with unknown dynamics. The system parameters and disturbances are drawn from unknown distributions, and the objective is to minimize the expected finite-horizon quadratic cost using observed trajectories. We establish a high-probability PAC--Bayes performance bound for a finite class of possibly time-varying nonlinear controllers. Under explicit polynomial-growth and moment conditions, the bound accommodates potentially unbounded quadratic costs and provides a theoretical guarantee on the expected control cost under unknown system uncertainties. Based on this bound, we develop a trajectory-based learning algorithm that evaluates candidate controllers using empirical quadratic costs and learns a posterior distribution over the controller class by minimizing the PAC--Bayes upper bound. The proposed framework does not require explicit identification of the unknown system dynamics and can accommodate nonlinear controllers, including neural networks. 
Numerical experiments on an omnidirectional mobile robot
demonstrate near-optimal performance without knowing
the true system parameters, improved performance over
a least-squares-based approach under parameter uncertainty,
and applicability to neural-network controllers.
\end{abstract}


\section{Introduction}\label{sec:introduction}

Nonlinear dynamical systems arise in many control applications,
including robotic manipulators, autonomous vehicles, and systems with
friction, actuator constraints, and uncertain physical parameters
\cite{kober2013robotics,mayne2000mpc}.
Their control often involves
nonconvex optimization and nonlinear and even non-differentiable closed-loop behavior. 
Such complexities can further increase the computational difficulty
of controller design, particularly in the presence of non-smooth dynamics~\cite{posa2014direct}.
Established methods for nonlinear systems such as iterative linear--quadratic regulation (iLQR),
iterative linear--quadratic Gaussian (iLQG) control, and nonlinear
model predictive control (MPC) address such problems using local
dynamics approximations or repeated finite-horizon non-convex optimisation
\cite{li2004ilqr,todorov2005ilqg,tassa2014ddp,mayne2000mpc}.
These approaches usually require a dynamics model or usable local
estimates of its derivatives. 
Other established approaches include feedback linearization,
backstepping, and robust and adaptive control
\cite{khalil2015nonlinear,kellett2023introduction}.
When the dynamics are uncertain, a model may instead be identified from data, for instance using sparse
regression or Gaussian processes
\cite{brunton2016sindy,deisenroth2015gp,hewing2020learningmpc}.
However, nonlinear model identification can itself be challenging,
often involving computationally demanding optimization problems,
such as nonlinear least squares~\cite{schoukens2019nonlinear}.
Direct data-driven controller synthesis provides another route,
although existing formulations often exploit particular system
structures~\cite{depersis2020formulas,pfrommer2023}.

We study a general data-driven setting: nonlinear systems whose
physical parameters vary across realizations according to a fixed but
unknown distribution. A motivating example is a family of products
manufactured by the same process. Products may exhibit variability in
mass, inertia, friction, or other parameters; related approaches such
as dynamics randomization also exploit variation in physical
parameters to improve transfer across systems~\cite{peng2018randomization}.
Precise parameter measurements can be costly and may require
disassembly or destructive testing in certain applications
\cite{madsen2020dynamics,douglas2002measurement}.
Instead, we consider
observed closed-loop state trajectories and the associated quadratic
costs. The aim is to learn a randomized controller-selection strategy
that performs well in expectation on previously unseen products
from the same population, without identifying each product's model
or observing its disturbances. This population-average objective is
distinct from worst-case robust or distributionally robust criteria
\cite{mohajerin2018dro}. With only finitely many rollout evaluations,
the discrepancy between empirical and population cost becomes central.

The probably approximately correct (PAC)--Bayes framework addresses
this generalization problem by upper-bounding the expected cost of a
randomized decision rule using its empirical cost and a
Kullback--Leibler (KL) penalty between a data-dependent posterior and
a data-independent prior
\cite{mcallester1999pac,mcallester2003pac,seeger2002pac,alquier2024pac}.
In control, \cite{majumdar2021pac} developed PAC--Bayes-based policy
learning with performance guarantees across previously unseen robotic
environments. More recently, \cite{boroujeni2024pac} proposed a
PAC--Bayesian approach to stochastic nonlinear optimal control with
stability guarantees. To apply bounded-cost PAC--Bayes results to an
unbounded objective, the latter work uses a $\tanh$ transformation of
the cost. PAC--Bayesian guarantees have also been investigated for
stochastic linear systems~\cite{eringis2021pac}; our earlier work
\cite{luo2026pac} considered unknown linear dynamics and unbounded
quadratic costs.

These developments motivate a PAC--Bayes bound tailored to
nonlinear systems with unknown random parameters and
polynomial-growth dynamics and controllers.
We establish an upper bound on the second moment of the potentially unbounded quadratic trajectory cost under polynomial-growth conditions and derive a high-probability PAC--Bayes guarantee.
The corresponding learning procedure evaluates a finite set of
candidate nonlinear controllers using repeated observed trajectories and
selects a posterior over these candidates without model identification.

\subsection{Contributions}
The main contributions of this paper are as follows.
\begin{itemize}
    \item We establish a high-probability PAC--Bayes performance bound
    for a finite class of possibly time-varying nonlinear controllers
    applied to nonlinear systems with unknown random parameters and
    disturbances. Specifically, the bound addresses the potentially unbounded quadratic cost under explicit growth and moment conditions.
    \item We develop a trajectory-based learning formulation algorithm based on the derived the PAC-Bayes bound that
    estimates the empirical quadratic cost of each candidate controller
    and learns a posterior distribution over the controller class.
    The resulting strategy targets expected control performance on
    unseen system realizations from the same distribution.
    \item Numerical experiments demonstrate that the proposed learning algorithm achieves competitive performance relative to the optimal benchmark, outperforms least-squares-based control under uncertain dynamics, and accommodates nonlinear neural-network controllers.
\end{itemize}

\subsection{Notation}
For a distribution $\mathcal D$, $\mathcal D^m$ denotes its $m$-fold
product distribution. The norm $\|x\|$ is Euclidean for vectors,
whereas $\|A\|$ and $\|A\|_F$ are the induced spectral and Frobenius
norms of a matrix. We use $\mathbb E$ and $\mathbb P$ for expectation
and probability. For probability
mass functions $\mathcal P,\mathcal P_0$ on
$\Pi=\{\pi^{(1)},\ldots,\pi^{(L)}\}$,
$
\mathrm{KL}(\mathcal P\Vert\mathcal P_0)
=\sum_{j=1}^{L}\mathcal P(\pi^{(j)})
\ln\frac{\mathcal P(\pi^{(j)})}{\mathcal P_0(\pi^{(j)})},
$
with $0\ln 0:=0$ and $\mathrm{KL}=+\infty$ if the posterior puts
positive mass outside the support of the prior. The Gamma function is
$\Gamma(z)=\int_0^\infty t^{z-1}e^{-t}\,dt$ for $z>0$. The notation $\operatorname{Beta}(a,b)$ denotes
the Beta distribution with density
$f(z)=z^{a-1}(1-z)^{b-1}/\mathrm{B}(a,b)$
for $0<z<1$, where
$\mathrm{B}(a,b)=\Gamma(a)\Gamma(b)/\Gamma(a+b)$.

\section{Problem Formulation}\label{sec:problem}

\subsection{Nonlinear Systems with Unknown Parameters}

Consider the finite-horizon discrete-time nonlinear system
\begin{equation}\label{eq:dynamics}
\begin{aligned}
    x_{t+1} &= f(x_t,u_t,w_t;\theta),\\
    u_t &= \pi_t(x_t),
    \qquad t=0,\ldots,T-1,
\end{aligned}
\end{equation}
where $x_t\in\mathbb R^{n_x}$ is the system state,
$u_t\in\mathbb R^{n_u}$ is the control input,
$w_t\in\mathbb R^{n_w}$ is the process disturbance, and $T$ is time horizon.
The control input is generated by a possibly nonlinear and
time-varying state-feedback controller
$\pi=(\pi_0,\ldots,\pi_{T-1})$, where
$\pi_t:\mathbb R^{n_x}\to\mathbb R^{n_u}$.
The unknown dynamics $f$ satisfy the conditions specified
in Section~\ref{subsec:assumption}.

The unknown system parameter $\theta$, initial state $x_0$ and process disturbances
$w_t$ satisfy
\begin{equation}
    \theta\sim\mathcal D_\theta,
    \quad x_0 \sim\mathcal D_0,\quad
    w_t\overset{\mathrm{i.i.d.}}{\sim}\mathcal D_w,
    \quad t=0,\ldots,T-1,
\end{equation}
where $\mathcal D_\theta$, $\mathcal D_0$ and $\mathcal D_w$ are fixed
but unknown distributions.
The disturbances are independent and identically distributed
across time, while $\theta$ is sampled once per trajectory
and remains fixed throughout the finite horizon.

The realizations of $\theta$ and $w_t$ are assumed to be unobservable,
whereas the initial state $x_0$ and the state trajectory
\[
    X_\pi=(x_0,x_1,\ldots,x_T)
\]
are observable.
The applied control inputs are also known since
$u_t=\pi_t(x_t)$.
The distribution $\mathcal P_{X_\pi}$ of $X_\pi$ is induced by
the unknown system parameters, initial states, and disturbances,
and depends on the chosen controller $\pi$.

\subsection{Assumptions}\label{subsec:assumption}

\begin{assumption}[System growth condition]
\label{ass:system}
There exist finite constants
$a_0,a_1,a_2,a_3\geq 0$ and exponents
$p_x,p_u,p_w>0$ such that, for
$\mathcal D_\theta$-almost every $\theta$,
\begin{equation}\label{eq:system_growth}
    \|f(x,u,w;\theta)\|
    \leq
    a_0+a_1\|x\|^{p_x}
    +a_2\|u\|^{p_u}
    +a_3\|w\|^{p_w}
\end{equation}
for all $(x,u,w)$.
The constants and exponents are independent of $\theta$.
\end{assumption}

Polynomial-growth conditions with different exponents
have been considered in nonlinear stochastic control
\cite{li2023discrete}.
Assumption~\ref{ass:system} adopts a similar approach
for discrete-time dynamics, allowing separate growth
exponents for the state, control input, and disturbance,
uniformly over the unknown parameters.
This condition is more general than global Lipschitz continuity,
since $\|f(x,u,w;\theta)\|\leq a_0+a_1\|x\|+a_2\|u\|+a_3\|w\|$
is a special case with $p_x=p_u=p_w=1$.

\begin{assumption}[Controller growth condition]
\label{ass:controller}
There exist finite constants $a_4,a_5\geq 0$
and an exponent $p_\pi>0$ such that
\begin{equation}\label{eq:controller_growth}
    \|\pi_t(x)\|
    \leq a_4+a_5\|x\|^{p_\pi},
    \qquad
    \forall x,\ \pi,\ 0\leq t<T.
\end{equation}
The constants are independent of the controller
$\pi$ and the time index $t$.
\end{assumption}

\begin{assumption}[Sub-Gaussian disturbances]
\label{ass:noise}
There exists a finite constant $\sigma_w>0$
such that each coordinate of the disturbance
satisfies
\begin{equation}\label{eq:subgaussian}
    \mathbb E\left[\exp(\eta w_{t,j})\right]
    \leq
    \exp\left(\frac{\eta^2\sigma_w^2}{2}\right),
    \qquad \forall\eta\in\mathbb R,
\end{equation}
where $w_{t,j}$ denotes the $j$-th coordinate of
$w_t$, for $j=1,\ldots,n_w$ and $t=0,\ldots,T-1$.
\end{assumption}

\subsection{Quadratic Cost and Learning Objective}
In this work, we consider the finite-horizon
quadratic cost
\begin{equation}
    \label{eq:J_pi_theta}
    J(\pi;\theta)
    :=\mathbb E_{x_0,W\mid\theta}
    \Bigg[
    \sum_{t=0}^{T-1}
    \left(
    x_t^\top Qx_t+u_t^\top Ru_t
    \right)
    +x_T^\top Qx_T
    \Bigg],
\end{equation}
where $W=(w_0,\ldots,w_{T-1})$,
$Q\succeq0$, and $R\succ0$ weight the state and
control input, respectively.
The expectation is taken over the random initial state
and disturbance sequence, conditional on $\theta$.

Since $\theta$ follows an unknown distribution,
we further define
\begin{equation}\label{eq:C_pi}
    C(\pi)
    :=\mathbb E_{\theta\sim\mathcal D_\theta}
    \left[J(\pi;\theta)\right],
\end{equation}
which can be equivalently expressed as
\begin{equation}\label{eq:C_over_Xpi}
    C(\pi)
    =
    \mathbb E_{X_\pi\sim \mathcal P_{X_\pi}}
    \left[C_q(\pi,X_\pi)\right],
\end{equation}
where the quadratic trajectory cost is
\begin{equation}
\label{eq:quadratic_cost}
    C_q(\pi,X_\pi)
    :=
    \sum_{t=0}^{T-1}
    \Bigl(
    x_t^\top Qx_t
    +\pi_t(x_t)^\top R\pi_t(x_t)
    \Bigr)
    +x_T^\top Qx_T.
\end{equation}
Here $\mathcal P_{X_\pi}$ denotes the probability distribution
of the state trajectory under controller $\pi$.

In this work, we consider a finite controller class
$\Pi=\{\pi^{(1)},\ldots,\pi^{(L)}\}$,
where each $\pi^{(j)}=(\pi_0^{(j)},\ldots,
\pi_{T-1}^{(j)})$ denotes a possibly time-varying
nonlinear controller.
A stochastic learning algorithm maps the observed
trajectory data to a data-dependent distribution
$\mathcal P$ over $\Pi$, from which a controller is sampled
according to $\pi\sim\mathcal P$.

The resulting expected cost is
\begin{equation}\label{eq:ultimate_cost}
    \bar C(\mathcal{P})
    :=
    \mathbb E_{\pi\sim \mathcal{P}}
    \left[C(\pi)\right].
\end{equation}
In the PAC--Bayes literature, $\bar C(\mathcal{P})$ is
referred to as the \emph{Gibbs expected cost}~\citep{c3}.

Our objective is to learn a controller from trajectory data
generated by unknown nonlinear systems.
We consider a stochastic learning algorithm that maps
the training dataset $\mathcal S$ to a data-dependent
distribution $\mathcal P$ over the finite controller
class $\Pi$, characterized by a probability mass function
satisfying
$    \mathcal P(\pi^{(j)})\geq 0,\quad
    \sum_{j=1}^{L}\mathcal P(\pi^{(j)})=1.$

A natural objective is to choose $\mathcal P$ to minimize
the Gibbs expected cost:
$    \min_{\mathcal P}\,\bar C(\mathcal P).$
However, since the nonlinear dynamics $f$ and the
distributions $\mathcal D_\theta$ and $\mathcal D_w$
are unknown, $\bar C(\mathcal P)$ cannot be
computed directly.

The PAC--Bayes approach serves two main purposes.
\begin{itemize}
    \item \textbf{Objective 1 (theoretical guarantees):}
    To derive a computable, data-dependent upper bound
    on $\bar C(\mathcal P)$ that provides
    high-probability performance guarantees
    for any posterior distribution $\mathcal P$,
    as discussed in Section~\ref{sec:theorem}.

    \item \textbf{Objective 2 (controller learning algorithm):}
    Based on the derived PAC--Bayes bound,
    to develop a data-driven learning algorithm
    that learns a distribution $\mathcal P$
    over $\Pi$ from finitely many observed trajectories,
    as discussed in Section~\ref{sec:algorithm}.
\end{itemize}

\section{PAC--Bayes Performance Guarantee}\label{sec:theorem}
In this section, we derive a high-probability PAC--Bayes
performance bound for a finite class of nonlinear controllers.
We first describe the data-generating process and then establish
a uniform upper bound on the second moment of the quadratic
trajectory cost under the polynomial-growth assumptions.
This leads to a performance guarantee for data-dependent
posterior distributions over controllers.
\subsection{Data-Generating Process}
For each controller $\pi^{(j)}\in\Pi$, $j=1,\ldots,L$, we
observe $n$ i.i.d. state trajectories
$X_{\pi^{(j)}}^{(1)},\ldots,X_{\pi^{(j)}}^{(n)}$
generated by \eqref{eq:dynamics}, with
$X_{\pi^{(j)}}^{(i)}\sim P_{X_{\pi^{(j)}}}$.
The joint law of the initial state, system parameter,
and disturbances is the same for every controller.
Datasets across different controllers may be dependent,
allowing the use of common random realizations to
facilitate controller comparisons.
Our analysis requires only i.i.d. trajectories
within each controller dataset.
Let
$\mathcal S=\{X_{\pi^{(j)}}^{(i)}:\,1\le j\le L,\ 1\le i\le n\}$.
The empirical cost of each controller and the Gibbs empirical cost are
\begin{align}
\widehat C(\pi^{(j)})
&:=\frac{1}{n}\sum_{i=1}^n
 C_q(\pi^{(j)},X_{\pi^{(j)}}^{(i)}),\label{eq:empirical_cost_nl}\\
\widetilde C(\mathcal P)
&:=\sum_{j=1}^{L}\mathcal P(\pi^{(j)})\widehat C(\pi^{(j)}).
\label{eq:gibbs_empirical_nl}
\end{align}
Both quantities are computed from observed trajectories and known
inputs; neither $\theta$ nor $w_t$ needs to be observed.

\subsection{PAC--Bayes Bound for a Finite Controller Class}
\begin{theorem}[PAC--Bayes bound for finite nonlinear controllers]
\label{thm:nonlinear_finite}
Consider \eqref{eq:dynamics} under
Assumptions~\ref{ass:system}--\ref{ass:noise},
with the quadratic cost defined in
\eqref{eq:quadratic_cost}.
Let $\mathcal P_0$ be any prior distribution on $\Pi$
fixed independently of $\mathcal S$, and let
$\Lambda\subset(0,\infty)$ be a nonempty finite set
fixed independently of $\mathcal S$.

Define
$ d:=\max\{1,p_x,p_up_\pi\},
 b:=\max\{1,p_\pi\}, h:=4bd^T.$
Suppose that the random initial state $x_0$ satisfies
$\mathbb E(1+\|x_0\|)^h\le M_0<\infty$.
For $t=0,\ldots,T$, define
$ q_t:=4bd^{T-t},
 r_0:=(M_0-1)^{1/h}.$
For $t=0,\ldots,T-1$, define
\begin{align*}
 r_{t+1}:={}&a_0+a_1r_t^{p_x}
 +2^{\max\{p_u-1,0\}}a_2
 \left(a_4^{p_u}+a_5^{p_u}r_t^{p_up_\pi}\right)\\
 &+a_3\min\!\left\{\vphantom{\frac{p_wq_{t+1}}2}\right.\\
 &\quad\left[2n_w(2n_w\sigma_w^2)^{p_wq_{t+1}/2}
 \Gamma\!\left(1+\frac{p_wq_{t+1}}2\right)\right]^{1/q_{t+1}},\\
 &\quad\left.\left[2n_w(2n_w\sigma_w^2)^{p_wh/2}
 \Gamma\!\left(1+\frac{p_wh}2\right)\right]^{1/h}
 \right\}.
\end{align*}
The controller-uniform second-moment constant is given by
\begin{equation}\label{eq:nonlinear_Bc}
 B_c:=\left[
 \|Q\|\sum_{t=0}^{T}r_t^2
 +\|R\|\sum_{t=0}^{T-1}
 \left(a_4+a_5 r_t^{p_\pi}\right)^2
 \right]^2.
\end{equation}
Then, for every $\delta\in(0,1)$, with probability
at least $1-\delta$ over $\mathcal S$, simultaneously
for all $\mathcal P\ll\mathcal P_0$ and $\lambda\in\Lambda$, we have
\begin{equation}\label{eq:pac_bayes_nonlinear}
 \bar C(\mathcal P)
 \le\widetilde C(\mathcal P)
 +\frac{\lambda B_c}{2n}
 +\frac{\mathrm{KL}(\mathcal P\Vert\mathcal P_0)
       +\ln(|\Lambda|/\delta)}{\lambda}.
\end{equation}
\end{theorem}

\begin{proof}[Proof sketch]
Since $h\ge1$ and $(1+z)^h\ge1+z^h$ for $z\ge0$,
\begin{equation*}
 \mathbb E\|x_0\|^h\le M_0-1=r_0^h.
\end{equation*}
Also, $q_0=h$ and
\begin{equation*}
 q_t=dq_{t+1}\ge p_xq_{t+1},\quad
 q_t\ge p_up_\pi q_{t+1},\quad
 q_t\ge4\max\{1,p_\pi\}.
\end{equation*}
For every $v>0$, the coordinatewise sub-Gaussian condition gives
\begin{equation*}
 \mathbb E\|w_t\|^v
 \le2n_w(2n_w\sigma_w^2)^{v/2}
 \Gamma\!\left(1+\frac v2\right).
\end{equation*}
Moreover, since $q_{t+1}\le h$, Lyapunov's inequality yields
\begin{align*}
 &\left(\mathbb E\|w_t\|^{p_wq_{t+1}}\right)^{1/q_{t+1}}\\
 &\quad\le \min\!\left\{
 \left[2n_w(2n_w\sigma_w^2)^{p_wq_{t+1}/2}
 \Gamma\!\left(1+\frac{p_wq_{t+1}}2\right)\right]^{1/q_{t+1}},\right.\\
 &\qquad\quad\left.
 \left[2n_w(2n_w\sigma_w^2)^{p_wh/2}
 \Gamma\!\left(1+\frac{p_wh}2\right)\right]^{1/h}
 \right\}.
\end{align*}
Assumptions~\ref{ass:system} and~\ref{ass:controller} imply
$ \|x_{t+1}\|
 \le a_0+a_1\|x_t\|^{p_x}
 +2^{\max\{p_u-1,0\}}a_2
 \bigl(a_4^{p_u}+a_5^{p_u}\|x_t\|^{p_up_\pi}\bigr)
 +a_3\|w_t\|^{p_w}.$
Taking $L^{q_{t+1}}$ norms, applying Minkowski's inequality
and the preceding moment-order inequalities, and inducting over $t$ give
\begin{equation*}
 \sup_{\pi\in\Pi}
 \left(\mathbb E\|x_t\|^{q_t}\right)^{1/q_t}
 \le r_t,\qquad t=0,\ldots,T.
\end{equation*}
As $q_t\ge4\max\{1,p_\pi\}$,
\begin{align*}
 \left(\mathbb E\|x_t\|^4\right)^{1/4}
 &\le r_t,\\
 \left(\mathbb E\|\pi_t(x_t)\|^4\right)^{1/4}
 &\le a_4+a_5 r_t^{p_\pi}.
\end{align*}
Minkowski's inequality for the nonnegative quadratic cost now gives
\begin{align*}
 &\left(\mathbb E C_q(\pi,X_\pi)^2\right)^{1/2}\\
 &\quad\le\|Q\|\sum_{t=0}^{T}
 \left(\mathbb E\|x_t\|^4\right)^{1/2}
 +\|R\|\sum_{t=0}^{T-1}
 \left(\mathbb E\|\pi_t(x_t)\|^4\right)^{1/2}\\
 &\quad\le \|Q\|\sum_{t=0}^{T}r_t^2
 +\|R\|\sum_{t=0}^{T-1}
 (a_4+a_5 r_t^{p_\pi})^2.
\end{align*}
Then, we have
\begin{equation*}\label{eq:uniform_second_moment}
 \sup_{\pi\in\Pi}\mathbb E\bigl[C_q(\pi,X_\pi)^2\bigr]
 \le B_c.
\end{equation*}

For fixed $\pi$, let $Z=C_q(\pi,X_\pi)\ge0$ and
$\mu=C(\pi)=\mathbb EZ$.
For every $\eta\ge0$, we have
\begin{equation*}
 1-\eta\mu+\frac{\eta^2}{2}\mathbb EZ^2
 \ge\frac12+\frac12(\eta\mu-1)^2>0.
\end{equation*}
Consequently, we get
\begin{align*}
 \log\mathbb E e^{\eta(\mu-Z)}
 &\le\eta\mu+
 \log\!\left(1-\eta\mu+\frac{\eta^2}{2}\mathbb EZ^2\right)\\
 &\le\frac{\eta^2}{2}\mathbb EZ^2
 \le\frac{\eta^2B_c}{2},
\end{align*}
where $e^{-z}\le1-z+z^2/2$ for $z\ge0$ is used.
The $n$ i.i.d. trajectories for each fixed controller imply
\begin{equation}\label{eq:one_sided_nl_mgf}
 \mathbb E_{\mathcal S}\exp\!\left(
 \lambda[C(\pi)-\widehat C(\pi)]\right)
 \le\exp\!\left(\frac{\lambda^2B_c}{2n}\right),
 \qquad\lambda>0.
\end{equation}
Since $\mathcal P_0$ is independent of $\mathcal S$,
Tonelli's theorem and~\eqref{eq:one_sided_nl_mgf} yield
\begin{equation*}
 \mathbb E_{\mathcal S}\int_\Pi
 \exp\!\left(\lambda[C(\pi)-\widehat C(\pi)]
 -\frac{\lambda^2B_c}{2n}\right)\mathcal P_0(d\pi)
 \le1.
\end{equation*}
Markov's inequality and a union bound over $\Lambda$ yield,
with probability at least $1-\delta$, for all $\lambda\in\Lambda$,
\begin{equation*}
 \log\int_\Pi\exp\!\left(
 \lambda[C(\pi)-\widehat C(\pi)]
 -\frac{\lambda^2B_c}{2n}\right)\mathcal P_0(d\pi)
 \le\ln\frac{|\Lambda|}{\delta}.
\end{equation*}
The Donsker--Varadhan change-of-measure inequality implies,
simultaneously for every $\mathcal P\ll\mathcal P_0$,
\begin{align*}
 \lambda\bigl[\bar C(\mathcal P)-\widetilde C(\mathcal P)\bigr]
 -\frac{\lambda^2B_c}{2n}
 \le{}&\mathrm{KL}(\mathcal P\Vert\mathcal P_0)
 +\ln\frac{|\Lambda|}{\delta}.
\end{align*}
Division by $\lambda>0$ completes~\eqref{eq:pac_bayes_nonlinear}. The complete proof is provided in Appendix~\ref{app:nonlinear_proof}.
\end{proof}

Theorem~\ref{thm:nonlinear_finite} provides a high-probability
upper bound on the expected quadratic cost for any posterior distribution over nonlinear controllers.
The bound applies to potentially unbounded costs without
requiring identification of the unknown system dynamics.
It balances the empirical cost against the KL divergence
from the prior, while the term $\lambda B_c/(2n)$
accounts for the uncertainty arising from finite data.
This provides a theoretical basis for learning nonlinear
controllers by using the PAC--Bayes bound.

\begin{remark}[Conservativeness of the bound]\label{remark:conservativeness}
Since our framework allows potentially unstable
systems, the uniform moment estimates account for
worst-case growth, which may lead to a conservative
constant $B_c$.
For certain stable nonlinear systems, however,
$B_c$ can remain moderate, yielding a more tight
PAC--Bayes bound, as illustrated in
Section~\ref{sec:experiments}.
Moreover, our approach preserves the original
nonlinear dynamics and quadratic cost, allowing
larger penalties for poorly performing controllers
and thereby facilitating their distinction from
better-performing ones.
\end{remark}

\section{PAC--Bayes Controller Learning Algorithms}
\label{sec:algorithm}

In this section, we address Objective~2 by learning a distribution over
controllers from observed trajectories. In view of
Theorem~\ref{thm:nonlinear_finite}, a key idea is to minimize the
PAC--Bayes upper bound. For each fixed $\lambda\in\Lambda$,
the terms $\lambda B_c/(2n)$ and $\ln(|\Lambda|/\delta)/\lambda$ do not
depend on the posterior. We therefore consider minimising the following problem
\begin{equation}\label{eq:learning_objective}
\mathcal P_\lambda\in\arg\min_{\mathcal P\ll\mathcal P_0}
\left\{\widetilde C(\mathcal P)
+\frac{\mathrm{KL}(\mathcal P\Vert\mathcal P_0)}{\lambda}\right\}.
\end{equation}
The theoretical bound remains valid for the learned posterior since it
holds simultaneously for all admissible posteriors and
all $\lambda\in\Lambda$.

\subsection{Finite-Controller Learning}
For $\Pi=\{\pi^{(1)},\ldots,\pi^{(L)}\}$, the empirical costs are
given by \eqref{eq:empirical_cost_nl}. For each fixed
$\lambda\in\Lambda$, Problem~\eqref{eq:learning_objective}
is convex and admits the following closed-form solution,
known as the Gibbs posterior~\cite{catoni2007pac,alquier2024pac}:
\begin{equation}\label{eq:optimal_posterior}
\mathcal P_\lambda(\pi^{(j)})=
\frac{\mathcal P_0(\pi^{(j)})
\exp[-\lambda\widehat C(\pi^{(j)})]}
{\sum_{k=1}^{L}\mathcal P_0(\pi^{(k)})
\exp[-\lambda\widehat C(\pi^{(k)})]}.
\end{equation}
We select $\lambda^\star$ by minimizing the
PAC--Bayes upper bound over $\Lambda$:
\begin{equation}\label{eq:optimal_lambda}
\lambda^\star\in\arg\min_{\lambda\in\Lambda}
\Bigg\{
\widetilde C(\mathcal P_\lambda)
+\frac{\lambda B_c}{2n}
+\frac{\mathrm{KL}(\mathcal P_\lambda\Vert\mathcal P_0)
+\ln(|\Lambda|/\delta)}{\lambda}
\Bigg\}.
\end{equation}
Here, $\lambda$ acts as a regularization parameter,
reflecting a trade-off between the tightness of
the PAC--Bayes bound and control performance.
The learned posterior is
$\mathcal P^\star=\mathcal P_{\lambda^\star}$.
The resulting procedure is summarized in
Algorithm~\ref{alg:finite_nonlinear}.

\begin{algorithm}[t]
\caption{PAC--Bayes Learning for Finite Controllers}
\label{alg:finite_nonlinear}
\begin{algorithmic}[1]
\Require Fixed $\Pi$, prior $\mathcal P_0$,
$n$ trajectories per controller,
finite grid $\Lambda$, $B_c$, $\delta\in(0,1)$.
\For{$j=1,\ldots,L$}
  \State Execute $\pi^{(j)}$ on $n$ independent trajectories.
  \State Compute $\widehat C(\pi^{(j)})$ by
  \eqref{eq:empirical_cost_nl}.
\EndFor
\For{$\lambda\in\Lambda$}
  \State Compute $\mathcal P_\lambda$ by
  \eqref{eq:optimal_posterior}.
  \State Evaluate the PAC--Bayes bound in
  \eqref{eq:pac_bayes_nonlinear}.
\EndFor
\State Select $\lambda^\star$ by \eqref{eq:optimal_lambda}.
\State Set $\mathcal P^\star=\mathcal P_{\lambda^\star}$.
\Ensure $\mathcal P^\star$; sample
$\pi\sim\mathcal P^\star$ for each new trajectory.
\end{algorithmic}
\end{algorithm}

\subsection{Neural-Network-Based Controller Learning}
A useful special case is a finite library of neural-network (NN)
controllers. For each $j=1,\ldots,L$ and $t=0,\ldots,T-1$, consider a
finite-depth feedforward network
\begin{align*}
 h_{0,t}^{(j)}&=x,\qquad \pi_t^{(j)}(x)=h_{D,t}^{(j)},\\
 h_{k,t}^{(j)}&=\sigma_k\!\left(W_{k,t}^{(j)}h_{k-1,t}^{(j)}
 +b_{k,t}^{(j)}\right),
\end{align*}
where $k=1,\ldots,D$, all weights and biases are finite, and each
activation $\sigma_k$ is $1$-Lipschitz and satisfies
$\sigma_k(0)=0$ (e.g., ReLU, $\tanh$, or the identity map).

\noindent\textit{Verification of Assumption~\ref{ass:controller}.}
Since each activation is $1$-Lipschitz and satisfies
$\sigma_k(0)=0$, we have
\begin{equation*}
\|h_{k,t}^{(j)}\|
\le\|W_{k,t}^{(j)}\|\,\|h_{k-1,t}^{(j)}\|
+\|b_{k,t}^{(j)}\|.
\end{equation*}
Applying this inequality recursively yields
\begin{align*}
 \|\pi_t^{(j)}(x)\|
 &\le a_4^{(j,t)}+a_5^{(j,t)}\|x\|,\\
 a_4^{(j,t)}
 &:=\sum_{k=1}^{D}
 \left(\prod_{r=k+1}^{D}\|W_{r,t}^{(j)}\|\right)
 \|b_{k,t}^{(j)}\|,\\
 a_5^{(j,t)}
 &:=\prod_{k=1}^{D}\|W_{k,t}^{(j)}\|.
\end{align*}
Since both $L$ and $T$ are finite, setting
$a_4=\max_{j,t}a_4^{(j,t)}$ and
$a_5=\max_{j,t}a_5^{(j,t)}$ verifies
\eqref{eq:controller_growth} with $p_\pi=1$.
Thus these NN controllers satisfy the same growth assumption and can
be handled by Theorem~\ref{thm:nonlinear_finite}.

Importantly, although the NN weights are required to take values from a finite set, the weights themselves can be real-valued. 
To this end, we consider $L$ preselected NNs with
fixed weights, which may be designed in advance
or pretrained using independent pilot data.
However, on contrast to the usual DNN training procedure where the weights are updated (usually via SGD) in training process, our PAC-Bayes learning does not update the (set of) NN weight values. 
Instead, only the posterior probabilities over
these fixed controllers are optimized using the
evaluation trajectories.
The Gibbs posterior and $\lambda^\star$ are obtained
from \eqref{eq:optimal_posterior} and
\eqref{eq:optimal_lambda}, respectively, following
Algorithm~\ref{alg:finite_nonlinear}.
Extending this approach to uncountable NN weight
spaces is left for future work.

\section{Numerical Experiments}\label{sec:experiments}

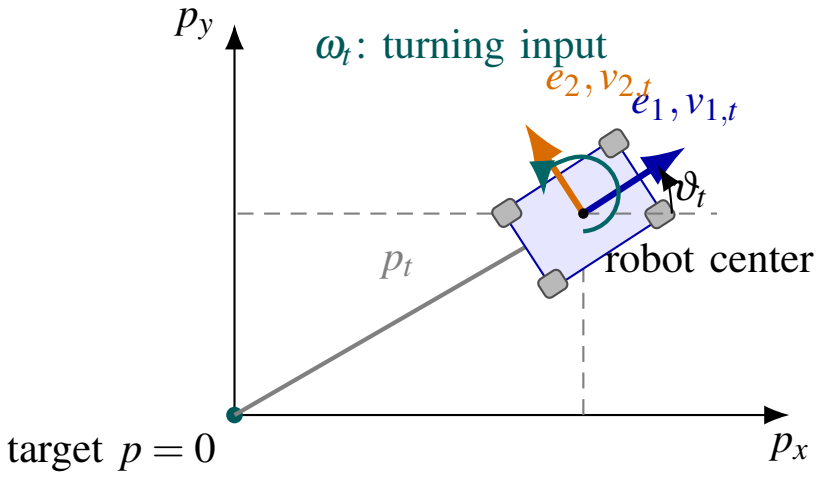
\begin{figure}[t]
\centering
\begin{tikzpicture}[scale=0.65,>=Latex,font=\footnotesize]
\draw[->] (0,0)--(5.65,0) node[below] {$p_x$};
\draw[->] (0,0)--(0,3.98) node[left] {$p_y$};
\fill[teal!70!black] (0,0) circle (2.4pt);
\node[below left] at (0,0) {target $p=0$};

\coordinate (C) at (3.55,2.05);
\draw[densely dashed,gray] (3.55,0)--(C)--(0,2.05);
\draw[->,thick,gray] (0,0)--(C)
node[pos=.58,above left] {$p_t$};

\begin{scope}[shift={(C)},rotate=33]
\filldraw[fill=blue!10,draw=blue!65!black,
rounded corners=2pt]
(-0.68,-0.47) rectangle (0.68,0.47);
\foreach \xx in {-0.65,0.65} {
\foreach \yy in {-0.43,0.43} {
\filldraw[fill=gray!55,draw=black!70,
rounded corners=1pt]
(\xx-0.13,\yy-0.11) rectangle
(\xx+0.13,\yy+0.11);
}}
\draw[->,very thick,blue!65!black]
(0,0)--(1.25,0) node[above] {$e_1,v_{1,t}$};
\draw[->,very thick,orange!85!black]
(0,0)--(0,1.10) node[above right] {$e_2,v_{2,t}$};
\draw[->,thick,teal!80!black]
(-0.1,-0.15) arc[start angle=-120,
end angle=150,radius=.38];
\end{scope}

\draw[dashed,gray] (C)--++(1.36,0);
\draw[->] ($(C)+(0.90,0)$)
arc[start angle=0,end angle=33,radius=.9];
\node at (4.64,2.30) {$\vartheta_t$};
\fill (C) circle (1.5pt);
\node[below right] at ($(C)+(0.03,-0.11)$)
{robot center};
\node[teal!70!black] at (2.3,3.73)
{$\omega_t$: turning input};
\end{tikzpicture}
\caption{\footnotesize Omnidirectional robot with body-frame
translational inputs $v_t$ and turning input $\omega_t$.
The target is $(p,\vartheta)=(0,0)$.}
\label{fig:robot}
\end{figure}

We consider an omnidirectional mobile robot, as
illustrated in Fig.~\ref{fig:robot}. The control objective
is to drive its position and heading towards zero while
minimizing the expected finite-horizon quadratic cost.
Since the dynamics are uncertain, we learn a controller
distribution by minimizing the corresponding PAC--Bayes
upper bound.
The state is $x_t=(p_t^\top,\vartheta_t)^\top$,
where $p_t=(p_{x,t},p_{y,t})^\top\in\mathbb R^2$ is the position and
$\vartheta_t$ is the heading angle. The control input
$u_t=(v_t^\top,\omega_t)^\top$ consists of the
body-frame translational velocity $v_t=(v_{1,t},v_{2,t})\in\mathbb R^2$
and angular velocity $\omega_t$.
The nonlinear dynamics are
\begin{align}
p_{t+1}
&=p_t+\theta_p\operatorname{Rot}(\vartheta_t)v_t+w_t^p,
\notag\\
\vartheta_{t+1}
&=\vartheta_t+\theta_\vartheta\omega_t+w_t^\vartheta,
\label{eq:robot_dynamics}
\end{align}
where $\operatorname{Rot}(\vartheta_t)$ is the planar
rotation matrix, and $\theta=(\theta_p,\theta_\vartheta)$
denotes the unknown actuation gains, capturing variations
in translational and rotational responses caused by
factors such as payload changes, ground friction,
and wheel slippage. We set
$Q=\operatorname{diag}(q_1,q_1,q_2)$ and
$R=\operatorname{diag}(r_1,r_1,r_2)$.

For known $\theta$, the globally optimal feedback
controller is
\begin{equation}\label{eq:robot_optimal}
\pi_t^{\mathrm{opt}}(x;\theta)=
\begin{bmatrix}
-k_{p,t}^{\mathrm{opt}}(\theta)
\operatorname{Rot}(\vartheta)^\top p\\
-k_{\vartheta,t}^{\mathrm{opt}}(\theta)\vartheta
\end{bmatrix},
\end{equation}
where the optimal feedback gains satisfy
\begin{align*}
k_{p,t}^{\mathrm{opt}}
&=\frac{\theta_p P_{t+1}}
{r_1+\theta_p^2P_{t+1}},&
P_t&=q_1+\frac{r_1P_{t+1}}
{r_1+\theta_p^2P_{t+1}},\\
k_{\vartheta,t}^{\mathrm{opt}}
&=\frac{\theta_\vartheta S_{t+1}}
{r_2+\theta_\vartheta^2S_{t+1}},&
S_t&=q_2+\frac{r_2S_{t+1}}
{r_2+\theta_\vartheta^2S_{t+1}},
\end{align*}
with $P_T=q_1$ and $S_T=q_2$.
The derivation of the optimal controller and the proof of global optimality are provided in Appendix~\ref{app:robot_optimal}.

For PAC--Bayes learning, we construct a finite
controller class using nominal feedback gains
computed with $\theta=(1,1)$:
\begin{equation*}
\pi_t^{(j)}(x)=
\begin{bmatrix}
-\alpha_p^{(j)}k_{p,t}^{\mathrm{nom}}
\operatorname{Rot}(\vartheta)^\top p\\
-\alpha_\vartheta^{(j)}
k_{\vartheta,t}^{\mathrm{nom}}\vartheta
\end{bmatrix}.
\end{equation*}

Both scaling factors are selected from
$\{0.5+0.875k/9:k=0,\ldots,9\}$, yielding
$L=100$ controllers with a uniform prior.
We set $(q_1,q_2,r_1,r_2)=(10,1,0.2,0.2)$,
$x_0=(2,-1.5,1.15)^\top$, $T=20$, and
$w_t\sim\mathcal N(0,0.09I_3)$.

\subsection{Comparison with the Optimal Controller}
We fix $\theta=(0.72,1.30)$ across all
trajectories and compare the PAC--Bayes method
with the globally optimal controller
\eqref{eq:robot_optimal} using the true parameters.
We consider $L=100$ predetermined controllers
and adopt a uniform prior
$\mathcal P_0(\pi^{(j)})=1/L$.
We use a fixed grid $\Lambda=\{1,2,3,4,5\}$
and set $\delta=0.05$. For each horizon $T\in\{20,30,\ldots,100\}$,
we generate $n=10$ independent training
trajectories per controller.
An additional $N_{\mathrm{test}}=100$ independent
test trajectories are used to estimate the
expected cost of each controller.
To compute $B_c$, we use the parameter bounds
$\theta_p,\theta_\vartheta\in[0.55,1.55]$.
Since $\operatorname{Rot}(\vartheta_t)$ is orthogonal,
the system and feedback controllers satisfy
Assumptions~\ref{ass:system} and
\ref{ass:controller}, respectively, with
the growth constants
$
a_0=0,\quad a_1=1,\quad a_2=1.55,\quad
a_3=1,\quad a_4=0,
$
$
a_5(T)=
\max_{\substack{j=1,\ldots,L\\t=0,\ldots,T-1}}
\left\{
|k_{p,t}^{(j)}|,
|k_{\vartheta,t}^{(j)}|
\right\},
$
and $p_x=p_u=p_w=p_\pi=1$.

\begin{figure}[t]
    \centering   \includegraphics[width=\columnwidth]{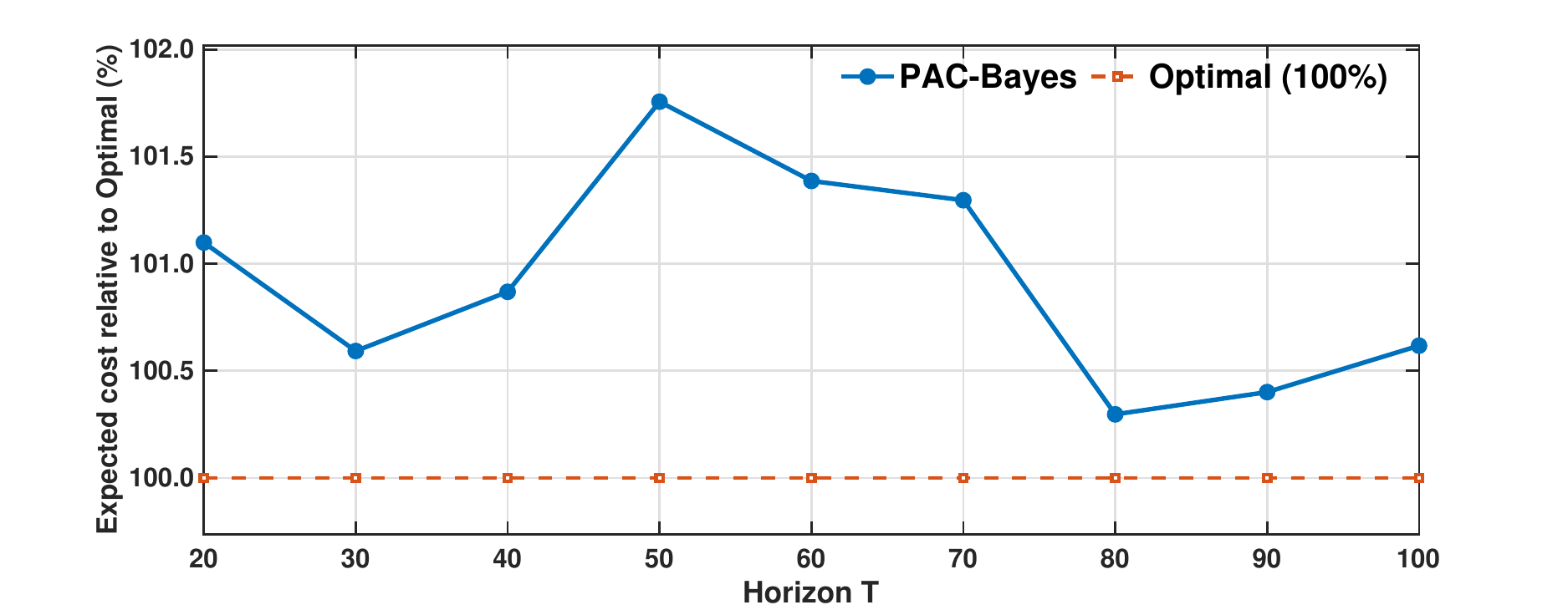}
    \caption{\footnotesize Normalized expected cost relative to the Bellman-optimal benchmark
versus horizon $T$ for fixed unknown parameters.
The PAC--Bayes method uses $n=10$ training trajectories per controller,
and both methods are evaluated using $N_{\mathrm{test}}=100$ held-out
disturbance trajectories. The Bellman-optimal cost is normalized to $100\%$.}
    \label{fig:robot_fixed_vs_T}
\end{figure}

Fig.~\ref{fig:robot_fixed_vs_T} compares the PAC--Bayes method with the globally optimal controller using the true parameters.
The empirical test costs are normalized by
the corresponding Bellman-optimal costs,
with the optimal benchmark fixed at $100\%$.
Across all horizons, the PAC--Bayes method
achieves near-optimal performance, with a
relative cost increase of less than $2\%$.
This demonstrates that the learned posterior achieves performance close to the optimal benchmark despite the unknown system parameters. 

\subsection{Comparison with Least-Squares-Based Control}
We draw the unknown system parameters independently for each trajectory from right-skewed distributions,
\begin{equation*}
\theta_p,\theta_\vartheta
\overset{\mathrm{i.i.d.}}{\sim}
0.55+\operatorname{Beta}(0.5,5),
\end{equation*}
and keep them fixed within each trajectory.
The underlying distributions and realized gains
are unknown to both compared methods.

We compare PAC--Bayes learning with a
least-squares (LS) identification baseline.
Specifically, Ceres Solver~\cite{agarwal2023ceres} is used to estimate
a single parameter vector
$\hat\theta_{\mathrm{LS}}$
by minimizing the pooled state-transition
squared errors over exactly the same observed
training trajectories used by PAC--Bayes.
The estimated parameters are then substituted
into \eqref{eq:robot_optimal}
to construct the LS-based controller.
We retain the same controller class, uniform prior,
finite grid $\Lambda$, and growth constants
as in the fixed-parameter setting.
For each horizon $T\in\{20,30,40,50\}$,
we use $n=100$ independent training trajectories
per controller and $N_{\mathrm{test}}=10000$
independent test trajectories.

\begin{figure}[t]
    \centering
    \includegraphics[width=\columnwidth]{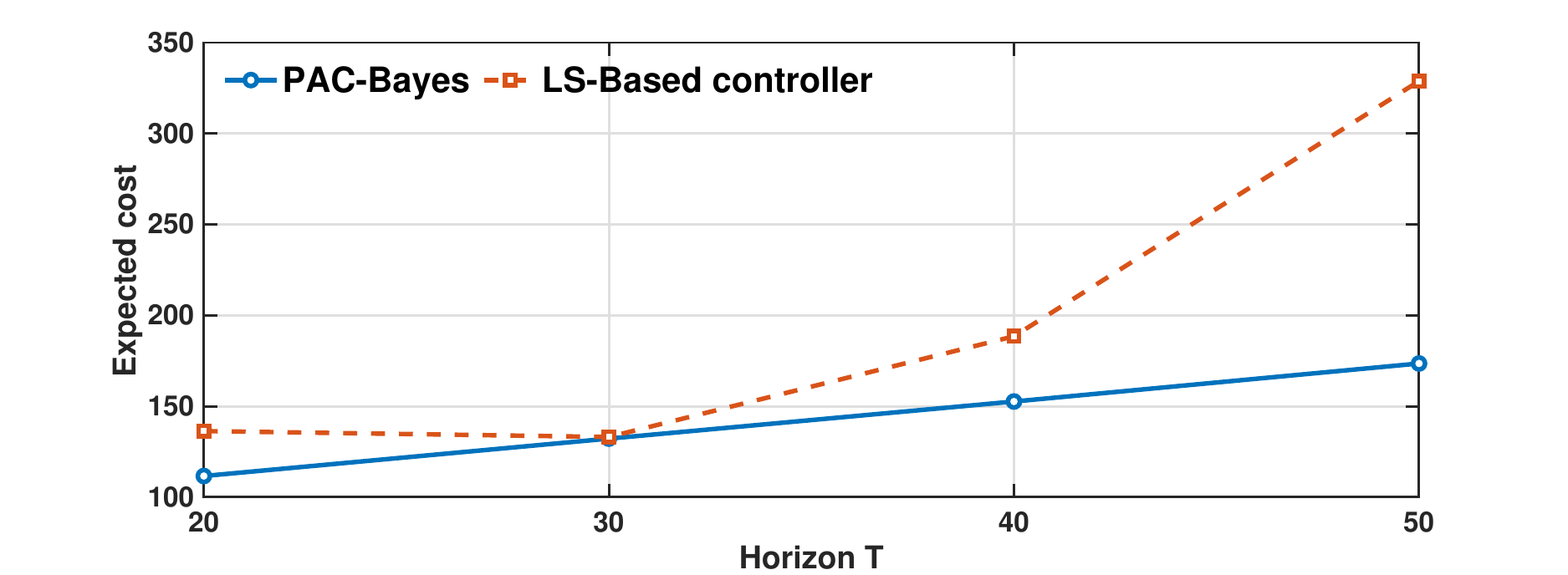}
    \caption{\footnotesize Test empirical quadratic cost versus horizon
    $T$ under random unknown parameters.
    PAC--Bayes method is compared with
    Ceres LS followed by Bellman control,
    using $n=100$ training trajectories per controller
    and $N_{\mathrm{test}}=10000$ test trajectories.}
    \label{fig:robot_random_vs_T}
\end{figure}

Fig.~\ref{fig:robot_random_vs_T} compares the
PAC--Bayes method with Ceres LS followed by
applying the optimal controller in
\eqref{eq:robot_optimal} using the estimated
system parameters, under randomly varying
system parameters.
The two methods achieve comparable test
empirical costs at $T=30$.
However, as the horizon increases, the
LS-based controller exhibits a substantially
larger increase in cost, whereas the
PAC--Bayes method maintains more gradual growth.
For the right-skewed parameter distribution,
LS tends to estimate an average parameter
value, which is more sensitive to rare,
relatively large parameter realizations.
In contrast, the median is less affected
by the upper tail and better represents
a typical realization.
Consequently, designing a controller using
a single LS parameter estimate may not
adequately account for the variability
of the system dynamics.
The PAC--Bayes method instead learns a
distribution over controllers directly
from trajectory costs, which may explain
its improved empirical performance in
this experiment.

\subsection{PAC--Bayesian Learning over a Finite Set of NN Controllers}
\noindent\textbf{NN Controller Library Construction.}
Following the construction in Section~\ref{sec:algorithm}, we first describe how we determine the set of controllers (i.e., library of NN). Specifically, we use $L=10$ neural networks (with their weights learned), constructed as below. The $j$th controller
has a hidden layer of 16 $\tanh$ units and an identity
output layer,
\begin{equation*}
\pi^{(j)}(x)=W_2^{(j)}\tanh(W_1^{(j)}x+b_1^{(j)})+b_2^{(j)}.
\end{equation*}
Each network is initially assigned an independent weight
initialization. Using independent pilot trajectories, we
optimize the mean $\mu_j$ of a Gaussian weight distribution
$q_j(W)=\mathcal N(\mu_j,\sigma_{\rm pilot}^2 I)$, with prior
$q_0(W)=\mathcal N(0,\sigma_0^2 I)$, by minimizing
\begin{equation*}
\mathbb E_{W\sim q_j}\mathbb E_{\xi}
\big[C(\pi_W,\xi)\big]
+\beta_{\rm pilot}\mathrm{KL}(q_j\|q_0).
\end{equation*}
The objective is optimized using antithetic zeroth-order
Gaussian gradient estimates and Adam, requiring only observed
pilot rollout costs and no derivatives of the unknown dynamics.
We subsequently use $W=\mu_j$ as the deterministic weights
of the $j$th controller and freeze all ten networks.
The pilot weight-distribution KL is distinct from the
controller-distribution KL used by Algorithm~\ref{alg:finite_nonlinear}.

\noindent\textbf{Verification of the Growth Condition.}
After the pilot training stage, all NN weights are
fixed, forming the finite controller class $\Pi$.
Since the hidden activation is bounded by one and
there are 16 hidden units, for every $x\in\mathbb R^3$,
$\|\pi^{(j)}(x)\|_2
\leq 4\|W_2^{(j)}\|_2+\|b_2^{(j)}\|_2.$
Hence, the frozen NN controllers satisfy
Assumption~\ref{ass:controller} with $p_\pi=1$,
$a_4=\max_{j=1,\ldots,L}
(4\|W_2^{(j)}\|_2+\|b_2^{(j)}\|_2)$,
and $a_5=0$.
These constants are then used to compute $B_c$
for the PAC--Bayes bound.

\noindent\textbf{PAC--Bayes Posterior Learning.}
With the controller library fixed, we apply
Algorithm~\ref{alg:finite_nonlinear} using
evaluation trajectories independent of the
pilot training data.
We use the uniform prior
$\mathcal P_0(\pi^{(j)})=1/10$
and $\Lambda=\{1,2,3,4,5\}$.
The algorithm optimizes only the posterior
probabilities over the fixed NN controllers
by minimizing the PAC--Bayes upper bound.
For $T=20$, we vary the number of evaluation
trajectories per controller over
$n\in\{10,20,50,100,200\}$,
with $\delta=0.05$.
We use an additional $N_{\mathrm{test}}=10^4$
independent trajectories, shared across
controllers as common test scenarios,
to estimate the expected costs.

\begin{figure}[t]
    \centering
    \includegraphics[width=\columnwidth]{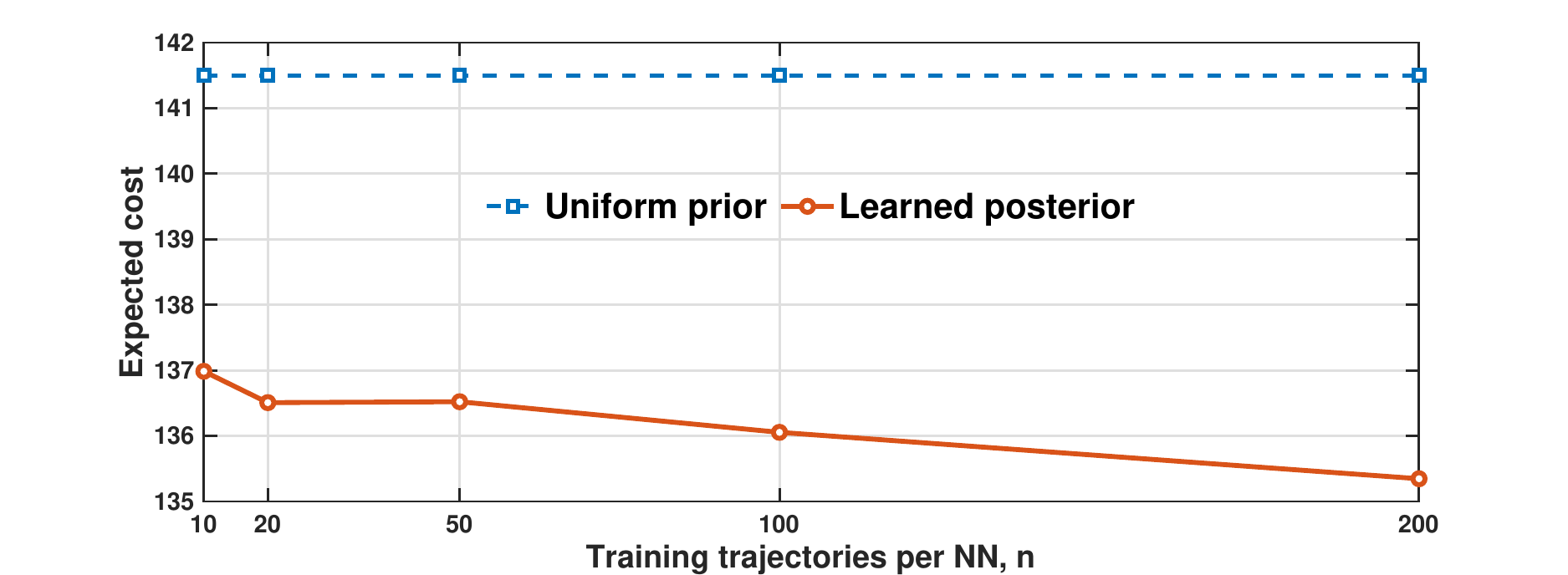}
    \caption{\footnotesize Comparison of the expected cost under the uniform prior and the learned
    PAC--Bayes posterior as the number of training
    trajectories per neural-network controller increases.}
    \label{fig:nn_prior_posterior}
\end{figure}
As shown in Fig.~\ref{fig:nn_prior_posterior},
the pretrained neural networks already provide
reasonably effective controllers, as reflected
by the performance under the uniform prior.
Nevertheless, PAC--Bayes learning can further
improve performance by learning a posterior
distribution that assigns greater probability
to better-performing controllers.
This demonstrates that the proposed approach
can complement neural-network-based control
by improving controller selection without
further updating the network weights.

For this robot model, the PAC--Bayes bound is
conservative due to the worst-case growth estimates,
which do not exploit the stability properties
of the closed-loop dynamics. We next illustrate that the bound can still
provide meaningful performance guarantees
for certain stable nonlinear systems.

\subsection{PAC--Bayes Bounds for Stable Nonlinear Systems}
\label{sec:stable_example}

As discussed in Remark~\ref{remark:conservativeness},
although the proposed PAC--Bayes bound can be
conservative, it can remain tight for certain
stable nonlinear systems. To illustrate this,
we consider
\begin{equation}\label{eq:stable_nonlinear_dynamics}
x_{t+1}=0.3x_t+0.15\tanh(2x_t)+\theta u_t+w_t,
\end{equation}
where $x_0=1$, $\theta\sim\mathcal U[0.9,1.1]$,
and $w_t\sim\mathcal N(0,0.1^2)$.
We set $T=6$, $Q=1$, $R=0.1$, and consider
$L=5$ feedback controllers
$\pi^{(j)}(x)=-k_jx$, where
$k_j\in\{0,0.05,0.10,0.15,0.20\}$.

The dynamics consist of a stable linear term
and a bounded nonlinear term.
Since
$|f(x,u,w;\theta)|
\leq 0.15+0.3|x|+1.1|u|+|w|$,
Assumptions~\ref{ass:system} and
\ref{ass:controller} hold with
$(a_0,a_1,a_2,a_3)=(0.15,0.3,1.1,1)$,
$(a_4,a_5)=(0,0.2)$, and all growth
exponents equal to one.
The corresponding moment recursion is
$r_0=15^{1/4}$ and
$r_{t+1}=0.35+0.52r_t$,
yielding $B_c\approx91.89$.
We apply Algorithm~\ref{alg:finite_nonlinear}
with a uniform prior,
$\Lambda=\{10^{-1+6\ell/200}:\ell=0,\ldots,200\}$,
and $n\in\{10,50,100,500,1000,5000,10000\}$
training trajectories per controller.
We set $\delta=0.05$, and use $10^5$
independent test trajectories to estimate
the expected cost.

\begin{figure}[t]
\centering
\includegraphics[width=\columnwidth]
{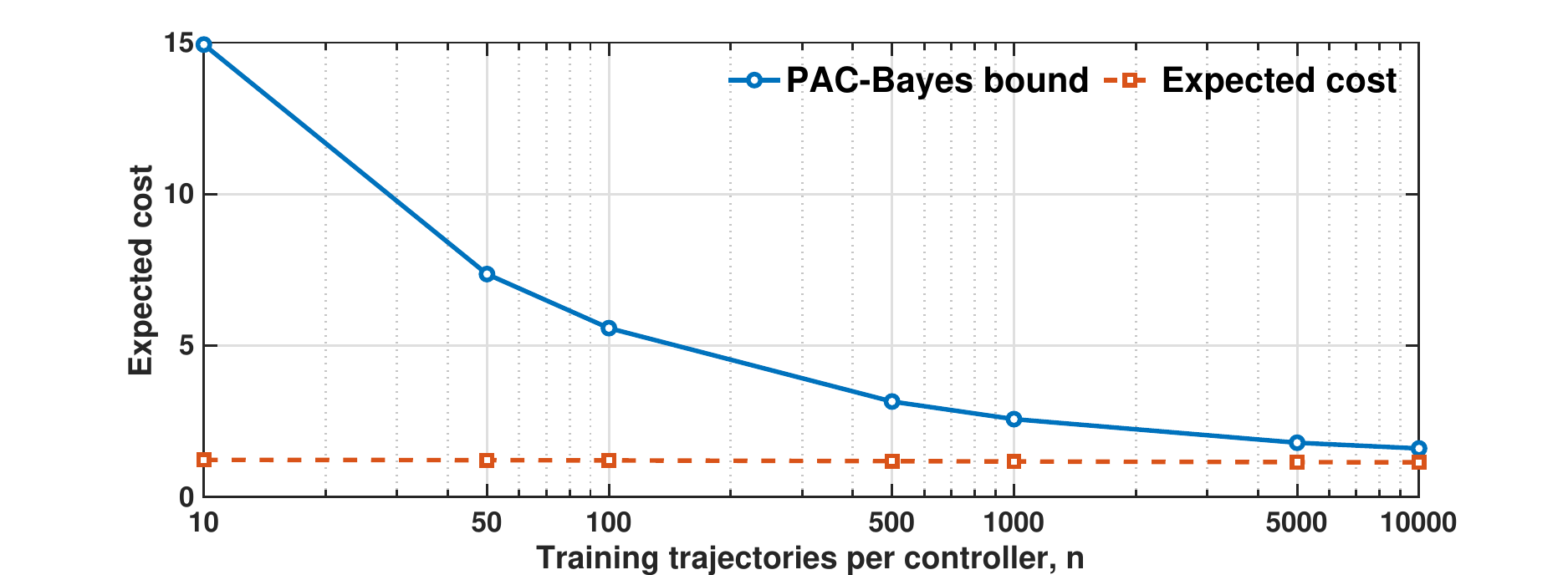}
\caption{\footnotesize PAC--Bayes performance bound and expected
cost for a stable nonlinear system with increasing
sample size.}
\label{fig:stable_bound}
\end{figure}

Fig.~\ref{fig:stable_bound} compares the
PAC--Bayes upper bound with the estimated
expected cost. As the sample size increases,
the bound decreases substantially and
gradually approaches the expected cost,
while the latter remains nearly constant.
This demonstrates that the bound can provide
a meaningful performance guarantee for
certain stable nonlinear systems.

\section{Conclusion and Future Work}
This paper established a PAC--Bayes performance bound for
nonlinear systems with unknown dynamics and developed a
trajectory-based controller learning algorithm.
Future work will focus on tightening the bound, extending
the framework to more complex nonlinear control tasks,
and deriving bounds for infinite controller classes to
enable more flexible neural-network controller learning.

\appendices
\section{Proof of Theorem~\ref{thm:nonlinear_finite}}
\label{app:nonlinear_proof}
Under the assumptions of Theorem~\ref{thm:nonlinear_finite},
$d=\max\{1,p_x,p_up_\pi\}$, $b=\max\{1,p_\pi\}$,
$h=4bd^T$, $q_t=4bd^{T-t}$, and $r_0=(M_0-1)^{1/h}$.
Define
\begin{equation*}
 M_w(v):=2n_w(2n_w\sigma_w^2)^{v/2}
 \Gamma\!\left(1+\frac v2\right),\qquad v>0.
\end{equation*}
For a nonnegative random variable $Y$ and $s>0$, write
\begin{equation*}
 \lVert Y\rVert_{L^s}:=(\mathbb E Y^s)^{1/s},\qquad s>0.
\end{equation*}
By Lyapunov's inequality,
\begin{equation*}
 0<s\le r\quad\Longrightarrow\quad
 \lVert Y\rVert_{L^s}\le\lVert Y\rVert_{L^r}
\end{equation*}
This holds also for $s<1$. Minkowski's inequality is applied only for $s\ge1$.

Fix a time index $t$, coordinate $j$, and threshold $z>0$.
For any $\eta>0$, Markov's inequality and
\eqref{eq:subgaussian} yield
\begin{align*}
 \mathbb P(w_{t,j}\ge z)
 &=\mathbb P(e^{\eta w_{t,j}}\ge e^{\eta z})\\
 &\le e^{-\eta z}\mathbb E e^{\eta w_{t,j}}\\
 &\le\exp\!\left(-\eta z+
 \frac{\eta^2\sigma_w^2}{2}\right).
\end{align*}
Minimizing the last exponent over $\eta>0$ gives
$\eta=z/\sigma_w^2$, hence
\begin{equation*}
 \mathbb P(w_{t,j}\ge z)\le e^{-z^2/(2\sigma_w^2)}.
\end{equation*}
Applying \eqref{eq:subgaussian} to $-w_{t,j}$ gives
\begin{equation*}
 \mathbb P(w_{t,j}\le-z)\le e^{-z^2/(2\sigma_w^2)}.
\end{equation*}
Combining the two tails,
\begin{equation*}
 \mathbb P(|w_{t,j}|\ge z)
 \le2e^{-z^2/(2\sigma_w^2)}.
\end{equation*}

If every coordinate satisfies $|w_{t,j}|<z/\sqrt{n_w}$,
then $\lVert w_t\rVert<z$. Therefore the union bound gives
\begin{align*}
 \mathbb P(\lVert w_t\rVert\ge z)
 &\le\sum_{j=1}^{n_w}
 \mathbb P \!\left(|w_{t,j}|\ge\frac z{\sqrt{n_w}}\right)
 \\
 &\le2n_w\exp\!\left(
 -\frac{z^2}{2n_w\sigma_w^2}\right).
\end{align*}
No independence among the coordinates of $w_t$ is needed here.

Let $v>0$. By Tonelli's theorem,
\begin{align*}
 \mathbb E \lVert w_t\rVert^v
 &=\mathbb E \int_0^{\lVert w_t\rVert}
 vz^{v-1}\,dz\\
 &=\int_0^\infty vz^{v-1}
 \mathbb P(\lVert w_t\rVert>z)\,dz.
\end{align*}
Using the vector tail bound,
\begin{equation*}
 \mathbb E \lVert w_t\rVert^v\le
 2n_wv\int_0^\infty z^{v-1}
 e^{-z^2/(2n_w\sigma_w^2)}\,dz.
\end{equation*}
Substituting $s=z^2/(2n_w\sigma_w^2)$ gives
\begin{equation*}
 \begin{aligned}
 z&=(2n_w\sigma_w^2)^{1/2}s^{1/2},\\
 dz&=\frac{(2n_w\sigma_w^2)^{1/2}}2
 s^{-1/2}\,ds,\\
 z^{v-1}dz&=\frac{(2n_w\sigma_w^2)^{v/2}}2
 s^{v/2-1}\,ds.
 \end{aligned}
\end{equation*}
Consequently,
\begin{align}
 \mathbb E \lVert w_t\rVert^v
 &\le n_wv(2n_w\sigma_w^2)^{v/2}
 \int_0^\infty s^{v/2-1}e^{-s}\,ds
 \notag\\
 &=n_wv(2n_w\sigma_w^2)^{v/2}
 \Gamma\!\left(\frac v2\right)\notag\\
 &=2n_w(2n_w\sigma_w^2)^{v/2}
 \Gamma\!\left(1+\frac v2\right)\notag\\
 &=M_w(v),\qquad v>0.
 \label{app:eq:complete_noise_moment}
\end{align}
Here we used $\Gamma(1+z)=z\Gamma(z)$.

By construction $d\ge1$ and $b\ge1$, so
\begin{equation*}
 h=q_0=4bd^T\ge4,
 \qquad q_t\ge q_T=4b\ge4.
\end{equation*}
Moreover,
\begin{equation*}
 q_t=dq_{t+1},\qquad t=0,\ldots,T-1.
\end{equation*}
The definition of $d$ yields
\begin{align*}
 p_xq_{t+1}&\le dq_{t+1}=q_t,
 \\
 p_up_\pi q_{t+1}&\le dq_{t+1}=q_t.
\end{align*}
Finally, because $b\ge p_\pi$ and $b\ge1$,
\begin{equation*}
 q_t\ge4b\ge4,\qquad q_t\ge4p_\pi.
\end{equation*}

The theorem assumes
\begin{equation*}
 \mathbb E (1+\lVert x_0\rVert)^h\le M_0<\infty.
\end{equation*}
Since $h\ge1$, for every $z\ge0$ one has
\begin{equation*}
 (1+z)^h\ge1+z^h.
\end{equation*}
Indeed, the function
$g(z)=(1+z)^h-z^h$ obeys $g(0)=1$ and
$g'(z)=h[(1+z)^{h-1}-z^{h-1}]\ge0$.
Taking expectations in the preceding scalar inequality
with $z=\lVert x_0\rVert$ gives
\begin{align*}
 1+\mathbb E \lVert x_0\rVert^h
 &\le\mathbb E (1+\lVert x_0\rVert)^h\\
 &\le M_0.
\end{align*}
As $q_0=h$ and $M_0\ge1$,
\begin{equation*}
 \lVert \lVert x_0\rVert\rVert_{L^{q_0}}
 \le(M_0-1)^{1/h}=r_0.
\end{equation*}

For any $p>0$ and $a,b\ge0$,
\begin{equation*}
 (a+b)^p\le2^{\max\{p-1,0\}}(a^p+b^p).
\end{equation*}
For $0<p\le1$ this follows from subadditivity of
$z\mapsto z^p$. For $p\ge1$ it follows from convexity:
$((a+b)/2)^p\le(a^p+b^p)/2$.

Applying the preceding inequality with
$p=p_u$, $a=a_4$, and $b=a_5\lVert x_t\rVert^{p_\pi}$,
\begin{align*}
 \lVert \pi_t(x_t)\rVert^{p_u}
 &\le(a_4+a_5\lVert x_t\rVert^{p_\pi})^{p_u}
 \\
 &\le2^{\max\{p_u-1,0\}}
 \bigl(a_4^{p_u}
 +a_5^{p_u}\lVert x_t\rVert^{p_up_\pi}\bigr).
\end{align*}
Thus, for $\mathcal D_\theta$-almost every $\theta$,
Assumptions~\ref{ass:system} and~\ref{ass:controller} imply
almost surely
\begin{equation}\label{app:eq:closed_loop_growth_detailed}
\begin{aligned}
 \lVert x_{t+1}\rVert
 &\le a_0+a_1\lVert x_t\rVert^{p_x}\\
 &\quad+2^{\max\{p_u-1,0\}}a_2
 \bigl(a_4^{p_u}
 +a_5^{p_u}\lVert x_t\rVert^{p_up_\pi}\bigr)\\
 &\quad+a_3\lVert w_t\rVert^{p_w}.
\end{aligned}
\end{equation}

We prove, for every $t=0,\ldots,T$,
\begin{equation}\label{app:eq:state_moment_induction_claim}
 \sup_{\pi\in\Pi}
 \lVert \lVert x_t\rVert\rVert_{L^{q_t}}\le r_t.
\end{equation}
The case $t=0$ follows from the initial moment bound.
Assume the estimate holds for a fixed $t<T$.
Fix any controller $\pi\in\Pi$ and put $s=q_{t+1}$.
From the definition of $q_t$, $s\ge4\ge1$.
Minkowski's inequality for the finite sum in
\eqref{app:eq:closed_loop_growth_detailed} gives
\begin{equation*}
\begin{aligned}
 \lVert \lVert x_{t+1}\rVert\rVert_{L^s}
 &\le a_0+a_1\lVert \lVert x_t\rVert^{p_x}\rVert_{L^s}\\
 &\quad+2^{\max\{p_u-1,0\}}a_2\Bigl[
 a_4^{p_u}\\
 &\hspace{43pt}+a_5^{p_u}
 \lVert \lVert x_t\rVert^{p_up_\pi}\rVert_{L^s}\Bigr]\\
 &\quad+a_3\lVert \lVert w_t\rVert^{p_w}\rVert_{L^s}.
\end{aligned}
\end{equation*}
For any $p>0$ and $s>0$,
\begin{equation*}
 \lVert Y^p\rVert_{L^s}=\lVert Y\rVert_{L^{ps}}^p
 \qquad(Y\ge0).
\end{equation*}
Combining the power identity with
Lyapunov's inequality, the moment-order inequalities, and the
induction hypothesis yields
\begin{align*}
 \lVert \lVert x_t\rVert^{p_x}\rVert_{L^s}
 &=\lVert \lVert x_t\rVert\rVert_{L^{p_xs}}^{p_x}
 \\
 &\le\lVert \lVert x_t\rVert\rVert_{L^{q_t}}^{p_x}
 \le r_t^{p_x},\\
 \lVert \lVert x_t\rVert^{p_up_\pi}\rVert_{L^s}
 &=\lVert \lVert x_t\rVert\rVert_{L^{p_up_\pi s}}^{p_up_\pi}
 \\
 &\le\lVert \lVert x_t\rVert\rVert_{L^{q_t}}^{p_up_\pi}
 \le r_t^{p_up_\pi}.
\end{align*}
Lyapunov's inequality also applies when $p_xs<1$ or $p_up_\pi s<1$.

For the disturbance term, \eqref{app:eq:complete_noise_moment} gives
\begin{align*}
 \lVert \lVert w_t\rVert^{p_w}\rVert_{L^s}
 &=\bigl(\mathbb E \lVert w_t\rVert^{p_ws}\bigr)^{1/s}
 \\
 &\le M_w(p_ws)^{1/s}.
\end{align*}
Since $s=q_{t+1}\le q_0=h$, Lyapunov's inequality also gives
\begin{align*}
 \lVert \lVert w_t\rVert^{p_w}\rVert_{L^s}
 &\le\lVert \lVert w_t\rVert^{p_w}\rVert_{L^h}
 \\
 &=\bigl(\mathbb E \lVert w_t\rVert^{p_wh}\bigr)^{1/h}
 \\
 &\le M_w(p_wh)^{1/h}.
\end{align*}
Hence,
\begin{equation*}
\begin{aligned}
 &\lVert \lVert w_t\rVert^{p_w}\rVert_{L^{q_{t+1}}}\\
 &\quad\le\min\!\left\{
 M_w(p_wq_{t+1})^{1/q_{t+1}},
 M_w(p_wh)^{1/h}\right\}.
\end{aligned}
\end{equation*}
Substituting these bounds into the preceding Minkowski inequality yields
\begin{align*}
 \lVert \lVert x_{t+1}\rVert\rVert_{L^{q_{t+1}}}
 &\le a_0+a_1r_t^{p_x}\\
 &\quad+2^{\max\{p_u-1,0\}}a_2
 (a_4^{p_u}+a_5^{p_u}r_t^{p_up_\pi})
 \\
 &\quad+a_3\min\!\left\{
 M_w(p_wq_{t+1})^{1/q_{t+1}},\right.\\
 &\hspace{66pt}\left.
 M_w(p_wh)^{1/h}\right\}
 \\
 &=r_{t+1}.
\end{align*}
Taking the supremum over $\Pi$ completes the induction. Since $r_0$, the growth constants, and $M_w(v)$ are finite, so is each $r_t$.

Since $4\le q_t$ by the moment-order inequalities,
Lyapunov's inequality and
\eqref{app:eq:state_moment_induction_claim} give
\begin{align*}
 \bigl(\mathbb E \lVert x_t\rVert^4\bigr)^{1/4}
 &=\lVert \lVert x_t\rVert\rVert_{L^4}\\
 &\le\lVert \lVert x_t\rVert\rVert_{L^{q_t}}
 \le r_t,\qquad 0\le t\le T.
\end{align*}
For $0\le t<T$, the controller-growth assumption yields
\begin{equation*}
 \lVert \pi_t(x_t)\rVert\le a_4+a_5\lVert x_t\rVert^{p_\pi}.
\end{equation*}
By Minkowski's inequality,
\begin{align*}
 \lVert \lVert \pi_t(x_t)\rVert\rVert_{L^4}
 &\le a_4+a_5\lVert \lVert x_t\rVert^{p_\pi}\rVert_{L^4}
 \\
 &=a_4+a_5\lVert \lVert x_t\rVert\rVert_{L^{4p_\pi}}^{p_\pi}
 \\
 &\le a_4+a_5\lVert \lVert x_t\rVert\rVert_{L^{q_t}}^{p_\pi}
 \\
 &\le a_4+a_5r_t^{p_\pi},
\end{align*}
where $4p_\pi\le q_t$. Both bounds are uniform over $\Pi$.

Fix $\pi$ and abbreviate
$Z_\pi:=C_q(\pi,X_\pi)$.
As $Q\succeq0$ and $R\succ0$,
\begin{equation*}
 \begin{aligned}
 0\le x^\top Qx&\le\lVert Q\rVert\lVert x\rVert^2,\\
 0\le u^\top Ru&\le\lVert R\rVert\lVert u\rVert^2.
 \end{aligned}
\end{equation*}
Hence \eqref{eq:quadratic_cost} and
the quadratic-form bounds above imply
\begin{align*}
 0\le Z_\pi
 &\le\lVert Q\rVert\sum_{t=0}^T\lVert x_t\rVert^2
 \\
 &\quad+\lVert R\rVert\sum_{t=0}^{T-1}
 \lVert \pi_t(x_t)\rVert^2.
\end{align*}
By Minkowski's inequality with exponent $2$,
\begin{equation*}
\begin{aligned}
 (\mathbb E Z_\pi^2)^{1/2}
 &\le\lVert Q\rVert\sum_{t=0}^T
 \lVert \lVert x_t\rVert^2\rVert_{L^2}\\
 &\quad+\lVert R\rVert\sum_{t=0}^{T-1}
 \lVert \lVert \pi_t(x_t)\rVert^2\rVert_{L^2}.
\end{aligned}
\end{equation*}
Since
\begin{equation*}
 \begin{aligned}
 \lVert \lVert x_t\rVert^2\rVert_{L^2}
 &=(\mathbb E \lVert x_t\rVert^4)^{1/2},\\
 \lVert \lVert \pi_t(x_t)\rVert^2\rVert_{L^2}
 &=(\mathbb E \lVert \pi_t(x_t)\rVert^4)^{1/2}.
 \end{aligned}
\end{equation*}
Using the fourth-moment estimates gives
\begin{equation*}
 \begin{aligned}
 (\mathbb E \lVert x_t\rVert^4)^{1/2}&\le r_t^2,\\
 (\mathbb E \lVert \pi_t(x_t)\rVert^4)^{1/2}
 &\le(a_4+a_5r_t^{p_\pi})^2.
 \end{aligned}
\end{equation*}
Substituting these bounds gives
\begin{equation*}
\begin{aligned}
 (\mathbb E Z_\pi^2)^{1/2}
 &\le\lVert Q\rVert\sum_{t=0}^Tr_t^2\\
 &\quad+\lVert R\rVert\sum_{t=0}^{T-1}
 (a_4+a_5r_t^{p_\pi})^2\\
 &=\sqrt{B_c}.
\end{aligned}
\end{equation*}
Therefore, for every $\pi\in\Pi$,
\begin{equation*}
 \mathbb E Z_\pi^2=\mathbb E[C_q(\pi,X_\pi)^2]\le B_c,
\end{equation*}
and so
\begin{equation*}
 \sup_{\pi\in\Pi}\mathbb E[C_q(\pi,X_\pi)^2]\le B_c.
\end{equation*}
In particular, $C(\pi)=\mathbb E Z_\pi$ is finite for every
controller because $\mathbb E Z_\pi\le(\mathbb E Z_\pi^2)^{1/2}$.

Fix a controller $\pi\in\Pi$ and write
\begin{equation*}
 Z:=C_q(\pi,X_\pi)\ge0,
 \qquad \mu:=\mathbb E Z=C(\pi).
\end{equation*}

For any real $z\ge0$,
\begin{equation*}
 e^{-z}\le1-z+\frac{z^2}{2}.
\end{equation*}
Indeed,
\begin{equation*}
 e^{-z}=1-z+\int_0^z(z-s)e^{-s}\,ds
 \le1-z+\int_0^z(z-s)\,ds.
\end{equation*}
Apply the preceding exponential inequality to
$z=\eta Z$ for arbitrary $\eta\ge0$. Taking
expectations yields
\begin{equation*}
 \mathbb E e^{-\eta Z}\le
 1-\eta\mu+\frac{\eta^2}{2}\mathbb E Z^2.
\end{equation*}
By Jensen's inequality,
$\mathbb E Z^2\ge(\mathbb E Z)^2=\mu^2$, and thus
\begin{align*}
 1-\eta\mu+\frac{\eta^2}{2}\mathbb E Z^2
 &\ge1-\eta\mu+\frac{(\eta\mu)^2}{2}
 \\
 &=\frac12+\frac12(\eta\mu-1)^2\\
 &\ge\frac12>0.
\end{align*}
Using
$\ln a\le a-1$ for every $a>0$,
\begin{align*}
 \ln\mathbb E e^{\eta(\mu-Z)}
 &=\eta\mu+\ln\mathbb E e^{-\eta Z}
 \\
 &\le\eta\mu+
 \ln\!\left(1-\eta\mu+
 \frac{\eta^2}{2}\mathbb E Z^2\right)
 \\
 &\le\eta\mu-\eta\mu
 +\frac{\eta^2}{2}\mathbb E Z^2
 \\
 &\le\frac{\eta^2B_c}{2}.
\end{align*}
Exponentiating,
\begin{equation*}
 \mathbb E \exp\bigl(\eta[C(\pi)-Z]\bigr)
 \le\exp\!\left(\frac{\eta^2B_c}{2}\right),
 \quad\eta\ge0.
\end{equation*}

For fixed $\pi=\pi^{(j)}$, define
\begin{equation*}
 Z_i:=C_q(\pi,X_\pi^{(i)}),
 \quad i=1,\ldots,n.
\end{equation*}
By the within-controller i.i.d. assumption,
$Z_1,\ldots,Z_n$ are i.i.d. with
$\mathbb E Z_i=C(\pi)=\mu$. Thus
\begin{equation*}
 \widehat C(\pi)=\frac1n\sum_{i=1}^nZ_i,
 \qquad
 \lambda[C(\pi)-\widehat C(\pi)]
 =\sum_{i=1}^n\frac\lambda n(\mu-Z_i).
\end{equation*}
For $\lambda>0$, independence within each controller dataset gives
\begin{align*}
 &\mathbb E_{\mathcal S}\exp\!\left(
 \lambda[C(\pi)-\widehat C(\pi)]\right)
 \\
 &\quad=\mathbb E \prod_{i=1}^n
 \exp\!\left(\frac\lambda n(\mu-Z_i)\right)
 \\
 &\quad=\prod_{i=1}^n
 \mathbb E \exp\!\left(\frac\lambda n(\mu-Z_i)\right)
 \\
 &\quad\le\prod_{i=1}^n
 \exp\!\left(\frac{\lambda^2B_c}{2n^2}\right)
 \\
 &\quad=\exp\!\left(\frac{\lambda^2B_c}{2n}\right).
\end{align*}
This proves \eqref{eq:one_sided_nl_mgf} without requiring independence across controller datasets.

Let $p_{0j}:=\mathcal P_0(\pi^{(j)})$.
For each $\lambda>0$, define
\begin{equation*}
\begin{aligned}
 H_\lambda(\mathcal S)
 :={}&\int_\Pi\exp\!\left(
 \lambda[C(\pi)-\widehat C(\pi)]
 -\frac{\lambda^2B_c}{2n}\right)
 \mathcal P_0(d\pi)\\
 ={}&\sum_{j=1}^Lp_{0j}
 \exp\!\left(
 \lambda[C(\pi^{(j)})-\widehat C(\pi^{(j)})]
 -\frac{\lambda^2B_c}{2n}\right).
\end{aligned}
\end{equation*}
By independence of $\mathcal P_0$ and $\mathcal S$, Tonelli's theorem, and \eqref{eq:one_sided_nl_mgf},
\begin{align*}
 \mathbb E_{\mathcal S}H_\lambda(\mathcal S)
 &=\sum_{j=1}^Lp_{0j}\mathbb E_{\mathcal S}
 \exp\!\left(
 \lambda[C(\pi^{(j)})-\widehat C(\pi^{(j)})]
 -\frac{\lambda^2B_c}{2n}\right)
 \\
 &\le\sum_{j=1}^Lp_{0j}=1.
\end{align*}

Fix $\delta\in(0,1)$ and the prespecified finite
grid $\Lambda$. For a fixed $\lambda\in\Lambda$,
Markov's inequality applied to $H_\lambda\ge0$ gives
\begin{align*}
 \mathbb P \!\left(
 H_\lambda(\mathcal S)>\frac{|\Lambda|}{\delta}
 \right)
 &\le\frac{\delta}{|\Lambda|}
 \mathbb E_{\mathcal S}H_\lambda(\mathcal S)
 \\
 &\le\frac{\delta}{|\Lambda|}.
\end{align*}
Taking a union bound over all $\lambda\in\Lambda$,
\begin{align*}
 &\mathbb P \!\left(
 \exists\lambda\in\Lambda:\
 H_\lambda(\mathcal S)>\frac{|\Lambda|}{\delta}
 \right)
 \\
 &\qquad\le\sum_{\lambda\in\Lambda}
 \frac\delta{|\Lambda|}=\delta.
\end{align*}
Therefore there exists an event $\mathcal E$,
with $\mathbb P(\mathcal E)\ge1-\delta$, such that
on $\mathcal E$, simultaneously for all
$\lambda\in\Lambda$,
\begin{equation}\label{app:eq:log_partition_event}
 \ln H_\lambda(\mathcal S)
 \le\ln\frac{|\Lambda|}{\delta}.
\end{equation}
The event $\mathcal E$ does not depend on the posterior.

Let $\mathcal P\ll \mathcal P_0$ be any posterior
on $\Pi$, and write
\begin{equation*}
 p_j:=\mathcal P(\pi^{(j)}),
 \qquad p_{0j}:=\mathcal P_0(\pi^{(j)}).
\end{equation*}
Absolute continuity means $p_{0j}=0$ implies
$p_j=0$. The finite-space KL divergence is
\begin{equation*}
 \mathrm{KL}(\mathcal P\Vert \mathcal P_0)
 =\sum_{j:p_j>0}p_j\ln\frac{p_j}{p_{0j}},
\end{equation*}
with the convention $0\ln 0=0$ for zero posterior
weights. On the support of $\mathcal P$ all denominators
$p_{0j}$ are strictly positive.

For arbitrary real numbers $g_1,\ldots,g_L$,
Jensen's inequality for concave $\ln$ gives
\begin{align*}
 &\sum_{j:p_j>0}p_j
 \ln\!\left(\frac{p_{0j}e^{g_j}}{p_j}\right)
 \\
 &\quad\le\ln\!\left(
 \sum_{j:p_j>0}p_j
 \frac{p_{0j}e^{g_j}}{p_j}\right)
 \\
 &\quad=\ln\!\left(
 \sum_{j:p_j>0}p_{0j}e^{g_j}\right)
 \\
 &\quad\le\ln\!\left(
 \sum_{j=1}^Lp_{0j}e^{g_j}\right).
\end{align*}
Expanding the left-hand side shows
\begin{equation}\label{app:eq:DV_finite}
 \sum_{j=1}^Lp_jg_j
 -\mathrm{KL}(\mathcal P\Vert \mathcal P_0)
 \le\ln\sum_{j=1}^Lp_{0j}e^{g_j}.
\end{equation}
This is the Donsker--Varadhan change-of-measure inequality.

On $\mathcal E$, for any $\lambda\in\Lambda$, set
\begin{equation*}
 g_j:=\lambda[C(\pi^{(j)})-
 \widehat C(\pi^{(j)})]
 -\frac{\lambda^2B_c}{2n}.
\end{equation*}
The right-hand side of \eqref{app:eq:DV_finite}
is exactly $\ln H_\lambda(\mathcal S)$.
By \eqref{app:eq:log_partition_event},
\begin{equation*}
 \sum_{j=1}^Lp_jg_j
 \le\mathrm{KL}(\mathcal P\Vert \mathcal P_0)
 +\ln\frac{|\Lambda|}{\delta}.
\end{equation*}
Expanding the left-hand side, using
$\sum_jp_j=1$, yields
\begin{align*}
 \sum_{j=1}^Lp_jg_j
 &=\lambda\sum_{j=1}^Lp_jC(\pi^{(j)})
 -\lambda\sum_{j=1}^Lp_j\widehat C(\pi^{(j)})
 -\frac{\lambda^2B_c}{2n}
 \\
 &=\lambda[\bar C(\mathcal P)
 -\widetilde C(\mathcal P)]
 -\frac{\lambda^2B_c}{2n}.
\end{align*}
Thus
\begin{equation*}
\begin{aligned}
 &\lambda[\bar C(\mathcal P)-\widetilde C(\mathcal P)]
 -\frac{\lambda^2B_c}{2n}\\
 &\qquad\le\mathrm{KL}(\mathcal P\Vert \mathcal P_0)
 +\ln\frac{|\Lambda|}{\delta}.
\end{aligned}
\end{equation*}
Rearranging and dividing by $\lambda>0$ gives
\begin{equation*}
\begin{aligned}
 \bar C(\mathcal P)
 &\le\widetilde C(\mathcal P)
 +\frac{\lambda B_c}{2n}\\
 &\quad+\frac{\mathrm{KL}(\mathcal P\Vert \mathcal P_0)
 +\ln(|\Lambda|/\delta)}{\lambda}.
\end{aligned}
\end{equation*}
Since \eqref{app:eq:DV_finite} holds for every $\mathcal P\ll\mathcal P_0$, and $\mathcal E$ holds simultaneously for all $\lambda\in\Lambda$, both $\mathcal P$ and $\lambda$ may be selected using $\mathcal S$. This proves~\eqref{eq:pac_bayes_nonlinear}.\hfill$\square$

\section{Derivation of the Globally Optimal Robot Controller}\label{app:robot_optimal}
\subsection{Bellman Equation}

Recall the state and input in Section~\ref{sec:experiments}:
\begin{align*}
p_t&=\begin{bmatrix}p_{x,t}\\p_{y,t}\end{bmatrix}
\in\mathbb R^2,\\
\vartheta_t&\in\mathbb R,\qquad
x_t=\begin{bmatrix}p_t\\\vartheta_t\end{bmatrix}
\in\mathbb R^3,\\
v_t&=\begin{bmatrix}v_{1,t}\\v_{2,t}\end{bmatrix}
\in\mathbb R^2,\\
\omega_t&\in\mathbb R,\qquad
u_t=\begin{bmatrix}v_t\\\omega_t\end{bmatrix}
\in\mathbb R^3.
\end{align*}
The heading $\vartheta_t\in\mathbb R$ is unwrapped.

For every $\vartheta\in\mathbb R$, let
\begin{equation*}
\operatorname{Rot}(\vartheta)=
\begin{bmatrix}
\cos\vartheta&-\sin\vartheta\\
\sin\vartheta&\cos\vartheta
\end{bmatrix}.
\end{equation*}
Taking the transpose gives
\begin{equation*}
\operatorname{Rot}(\vartheta)^\top=
\begin{bmatrix}
\cos\vartheta&\sin\vartheta\\
-\sin\vartheta&\cos\vartheta
\end{bmatrix}.
\end{equation*}
By direct calculation,
\begin{align*}
&\operatorname{Rot}(\vartheta)^\top\operatorname{Rot}(\vartheta)\\
&=\begin{bmatrix}
\cos^2\vartheta+\sin^2\vartheta &
-\cos\vartheta\sin\vartheta+\sin\vartheta\cos\vartheta\\
-\sin\vartheta\cos\vartheta+\cos\vartheta\sin\vartheta &
\sin^2\vartheta+\cos^2\vartheta
\end{bmatrix}\\
&=\begin{bmatrix}1&0\\0&1\end{bmatrix}=I_2.
\end{align*}
Similarly,
\begin{equation*}
\operatorname{Rot}(\vartheta)\operatorname{Rot}(\vartheta)^\top=I_2.
\end{equation*}
Consequently,
\begin{align*}
\|\operatorname{Rot}(\vartheta)z\|^2
&=z^\top\operatorname{Rot}(\vartheta)^\top\operatorname{Rot}(\vartheta)z\\
&=z^\top z=\|z\|^2,\\
\|\operatorname{Rot}(\vartheta)^\top z\|^2
&=z^\top\operatorname{Rot}(\vartheta)\operatorname{Rot}(\vartheta)^\top z\\
&=\|z\|^2
\end{align*}
for every $z\in\mathbb R^2$.

Recall the dynamics~\eqref{eq:robot_dynamics}, for $t=0,\ldots,T-1$:
\begin{align*}
p_{t+1}
&=p_t+\theta_p\operatorname{Rot}(\vartheta_t)v_t+w_t^p,
\\
\vartheta_{t+1}
&=\vartheta_t+\theta_\vartheta\omega_t+w_t^\vartheta.
\end{align*}

The disturbances are
\begin{equation*}
w_t^p=\begin{bmatrix}w_{1,t}^p\\w_{2,t}^p\end{bmatrix},
\quad w_t^\vartheta\in\mathbb R,
\quad w_t=\begin{bmatrix}w_t^p\\w_t^\vartheta\end{bmatrix}.
\end{equation*}

The translational equation in matrix form is
\begin{align*}
\begin{bmatrix}p_{x,t+1}\\p_{y,t+1}\end{bmatrix}
={}&\begin{bmatrix}p_{x,t}\\p_{y,t}\end{bmatrix}\\
&+\theta_p\begin{bmatrix}
\cos\vartheta_t&-\sin\vartheta_t\\
\sin\vartheta_t&\cos\vartheta_t
\end{bmatrix}\begin{bmatrix}v_{1,t}\\v_{2,t}\end{bmatrix}
+\begin{bmatrix}w_{1,t}^p\\w_{2,t}^p\end{bmatrix}.
\end{align*}
Equivalently,
\begin{align*}
p_{x,t+1}
={}&p_{x,t}+\theta_p v_{1,t}\cos\vartheta_t
-\theta_p v_{2,t}\sin\vartheta_t+w_{1,t}^p,\\
p_{y,t+1}
={}&p_{y,t}+\theta_p v_{1,t}\sin\vartheta_t
+\theta_p v_{2,t}\cos\vartheta_t+w_{2,t}^p,\\
\vartheta_{t+1}
={}&\vartheta_t+\theta_\vartheta\omega_t+w_t^\vartheta.
\end{align*}

Let $\mathcal F_t$ denote the information available immediately before the time-$t$ control is selected, including the current state and the known gains. For the Bellman benchmark, assume the noise is independent of $\mathcal F_t$ and of the selected input, is zero-mean, and has finite time-invariant second moments:
\begin{align*}
\mathbb E[w_t^p\mid\mathcal F_t]&=0,
&\mathbb E[w_t^\vartheta\mid\mathcal F_t]&=0,
\\
\mathbb E[w_t^p(w_t^p)^\top\mid\mathcal F_t]
&=\Sigma_p, &
\mathbb E[(w_t^\vartheta)^2\mid\mathcal F_t]
&=\sigma_\vartheta^2,
\end{align*}
where $\Sigma_p\in\mathbb R^{2\times2}$ is finite, symmetric, and positive semidefinite and $\sigma_\vartheta^2<\infty$. These conditional moments hold for the independent Gaussian disturbances used in Section~\ref{sec:experiments}, $w_t\sim\mathcal N(0,0.09I_3)$, in which case $\Sigma_p=0.09I_2$ and $\sigma_\vartheta^2=0.09$. No zero cross-covariance between $w_t^p$ and $w_t^\vartheta$ is needed for the following derivation.

Write
\begin{equation*}
\Sigma_p=\begin{bmatrix}
\sigma_{11}&\sigma_{12}\\
\sigma_{12}&\sigma_{22}
\end{bmatrix},\qquad
\operatorname{tr}(\Sigma_p)=\sigma_{11}+\sigma_{22}.
\end{equation*}
If the two translational noises have variances $\sigma_{p,1}^2$ and $\sigma_{p,2}^2$, then $\operatorname{tr}(\Sigma_p)=\sigma_{p,1}^2+\sigma_{p,2}^2$.

Assume
\begin{equation*}
\mathbb E[\|x_0\|^2\mid\theta]<\infty,
\qquad q_1>0,\;q_2>0,\;r_1>0,\;r_2>0.
\end{equation*}
The derivation is conditional on fixed, known $\theta=(\theta_p,\theta_\vartheta)$, as in the optimal benchmark \eqref{eq:robot_optimal}. Controls range over $\mathbb R^2\times\mathbb R$. No sign restriction on the gains is needed.

The state and control weights are
\begin{equation*}
Q=\begin{bmatrix}q_1&0&0\\0&q_1&0\\0&0&q_2\end{bmatrix},
\qquad
R=\begin{bmatrix}r_1&0&0\\0&r_1&0\\0&0&r_2\end{bmatrix}.
\end{equation*}
Then
\begin{align*}
x_t^\top Qx_t
&=\begin{bmatrix}p_{x,t}&p_{y,t}&\vartheta_t\end{bmatrix}
Q\begin{bmatrix}p_{x,t}\\p_{y,t}\\\vartheta_t\end{bmatrix}\\
&=q_1p_{x,t}^2+q_1p_{y,t}^2+q_2\vartheta_t^2\\
&=q_1\|p_t\|^2+q_2\vartheta_t^2,
\end{align*}
\begin{align*}
u_t^\top Ru_t
&=\begin{bmatrix}v_{1,t}&v_{2,t}&\omega_t\end{bmatrix}
R\begin{bmatrix}v_{1,t}\\v_{2,t}\\\omega_t\end{bmatrix}\\
&=r_1v_{1,t}^2+r_1v_{2,t}^2+r_2\omega_t^2\\
&=r_1\|v_t\|^2+r_2\omega_t^2.
\end{align*}

For a causal policy $\pi=(\pi_0,\ldots,\pi_{T-1})$, the conditional cost is
\begin{align*}
J(\pi;\theta)
={}&\mathbb E_{x_0,W\mid\theta}\Bigg[\sum_{t=0}^{T-1}
\bigl(x_t^\top Qx_t+u_t^\top Ru_t\bigr)
+x_T^\top Qx_T\Bigg]
\\
={}&\mathbb E_{x_0,W\mid\theta}\Bigg[\sum_{t=0}^{T-1}
\Bigl(q_1\|p_t\|^2+q_2\vartheta_t^2
+r_1\|v_t\|^2+r_2\omega_t^2\Bigr)
\\
&\hspace{13mm}+q_1\|p_T\|^2+q_2\vartheta_T^2\Bigg].
\end{align*}
Expanding the cost gives
\begin{align*}
\sum_{t=0}^{T-1}\bigl(&q_1p_{x,t}^2+q_1p_{y,t}^2+q_2\vartheta_t^2
+r_1v_{1,t}^2\\
&+r_1v_{2,t}^2+r_2\omega_t^2\bigr)
+q_1p_{x,T}^2+q_1p_{y,T}^2+q_2\vartheta_T^2.
\end{align*}
The optimal-control objective is
\begin{equation*}
\pi^{\mathrm{opt}}\in\operatorname*{arg\,min}_{\pi}J(\pi;\theta).
\end{equation*}

For fixed gains, let $V_t(p ,\vartheta)$ denote the smallest conditional expected remaining cost from time $t$ through the terminal time $T$, given the current state $(p ,\vartheta)$. By the principle of optimality,
\begin{align*}
V_T(p ,\vartheta)
&=q_1\|p \|^2+q_2\vartheta^2,
\\
V_t(p ,\vartheta)
&=\min_{v \in\mathbb R^2,\,\omega\in\mathbb R}
\Bigl\{q_1\|p \|^2+q_2\vartheta^2
+r_1\|v \|^2+r_2\omega^2
\\[-2pt]
&\hspace{4mm}+\mathbb E \bigl[V_{t+1}(p^+,\vartheta^+)\bigr]\Bigr\},
\end{align*}
where the next-state variables are
\begin{align*}
p^+&=p +\theta_p\operatorname{Rot}(\vartheta)v +w_t^p,\\
\vartheta^+&=\vartheta+\theta_\vartheta\omega+w_t^\vartheta.
\end{align*}

In the Bellman equation, expectations are conditional on the current state, action, and $\theta$.

\subsection{Quadratic Value Function}
At $t=T$,
\begin{align*}
V_T(p ,\vartheta)
&=\begin{bmatrix}p^\top&\vartheta\end{bmatrix}
\begin{bmatrix}q_1I_2&0\\0&q_2\end{bmatrix}
\begin{bmatrix}p \\\vartheta\end{bmatrix}\\
&=q_1p ^\top p +q_2\vartheta^2\\
&=q_1\|p \|^2+q_2\vartheta^2\\
&=P_T\|p \|^2+S_T\vartheta^2+c_T,
\end{align*}
with the terminal coefficients
\begin{equation*}
{P_T=q_1,\qquad S_T=q_2,\qquad c_T=0.}
\end{equation*}

For a fixed $t\in\{0,\ldots,T-1\}$, suppose
\begin{equation*}
V_{t+1}(p ,\vartheta)
=P_{t+1}\|p \|^2
+S_{t+1}\vartheta^2+c_{t+1},
\end{equation*}
where $P_{t+1},S_{t+1},c_{t+1}\ge0$ depend only on the fixed gains, time index, and cost and noise parameters.

Substituting the next state gives
\begin{align*}
V_{t+1}(p^+,\vartheta^+)
={}&P_{t+1}\|p^+\|^2
+S_{t+1}(\vartheta^+)^2+c_{t+1}\\
={}&P_{t+1}\|p +\theta_p\operatorname{Rot}(\vartheta)v
+w_t^p\|^2\\
&+S_{t+1}(\vartheta+\theta_\vartheta\omega
+w_t^\vartheta)^2+c_{t+1}.
\end{align*}
The Bellman equation becomes
\begin{align*}
V_t(p ,\vartheta)
=\min_{v ,\omega}\Bigl\{&q_1\|p \|^2+q_2\vartheta^2
+r_1\|v \|^2+r_2\omega^2
\\
&+P_{t+1}\mathbb E \|p +\theta_p\operatorname{Rot}(\vartheta)v
+w_t^p\|^2
\\
&+S_{t+1}\mathbb E (\vartheta+\theta_\vartheta\omega
+w_t^\vartheta)^2+c_{t+1}\Bigr\}.
\end{align*}

Let $a=p+\theta_p\operatorname{Rot}(\vartheta)v$. Then
\begin{align*}
\|a+w_t^p\|^2
&=(a+w_t^p)^\top(a+w_t^p)\\
&=a^\top a+a^\top w_t^p
+(w_t^p)^\top a
+(w_t^p)^\top w_t^p\\
&=\|a\|^2+2a^\top w_t^p
+\|w_t^p\|^2.
\end{align*}
Taking conditional expectations gives
\begin{align*}
\mathbb E \|a+w_t^p\|^2
={}&\mathbb E \|a\|^2+2\mathbb E[a^\top w_t^p]
+\mathbb E \|w_t^p\|^2\\
={}&\|a\|^2+2a^\top\mathbb E[w_t^p]
+\mathbb E[(w_t^p)^\top w_t^p]\\
={}&\|a\|^2+\mathbb E[(w_t^p)^\top w_t^p].
\end{align*}
By the trace identity,
\begin{align*}
\mathbb E[(w_t^p)^\top w_t^p]
&=\mathbb E \operatorname{tr}\bigl((w_t^p)^\top w_t^p\bigr)\\
&=\mathbb E \operatorname{tr}\bigl(w_t^p(w_t^p)^\top\bigr)\\
&=\operatorname{tr}\mathbb E \bigl[w_t^p(w_t^p)^\top\bigr]\\
&=\operatorname{tr}(\Sigma_p).
\end{align*}
Equivalently,
\begin{align*}
\mathbb E \|w_t^p\|^2
&=\mathbb E[(w_{1,t}^p)^2+(w_{2,t}^p)^2]\\
&=\mathbb E[(w_{1,t}^p)^2]+\mathbb E[(w_{2,t}^p)^2]\\
&=(\Sigma_p)_{11}+(\Sigma_p)_{22}\\
&=\operatorname{tr}(\Sigma_p).
\end{align*}
Hence,
\begin{equation*}
{\begin{aligned}
&\mathbb E \|p +\theta_p\operatorname{Rot}(\vartheta)v +w_t^p\|^2\\
&\quad=\|p +\theta_p\operatorname{Rot}(\vartheta)v \|^2
+\operatorname{tr}(\Sigma_p).
\end{aligned}}
\end{equation*}

For the angular component,
\begin{align*}
&(\vartheta+\theta_\vartheta\omega+w_t^\vartheta)^2\\
&=\bigl((\vartheta+\theta_\vartheta\omega)
+w_t^\vartheta\bigr)^2\\
&=(\vartheta+\theta_\vartheta\omega)^2
+2(\vartheta+\theta_\vartheta\omega)w_t^\vartheta
+(w_t^\vartheta)^2.
\end{align*}
Taking expectations gives
\begin{align*}
&\mathbb E (\vartheta+\theta_\vartheta\omega+w_t^\vartheta)^2\\
&=(\vartheta+\theta_\vartheta\omega)^2
+2(\vartheta+\theta_\vartheta\omega)\mathbb E[w_t^\vartheta]
+\mathbb E[(w_t^\vartheta)^2]\\
&=(\vartheta+\theta_\vartheta\omega)^2
+0+\sigma_\vartheta^2.
\end{align*}
Therefore,
\begin{equation*}
{\mathbb E (\vartheta+\theta_\vartheta\omega+w_t^\vartheta)^2
=(\vartheta+\theta_\vartheta\omega)^2+\sigma_\vartheta^2.}
\end{equation*}

Substituting the two conditional expectations gives
\begin{align*}
V_t(p ,\vartheta)
=\min_{v ,\omega}\Bigl\{&q_1\|p \|^2+q_2\vartheta^2
+r_1\|v \|^2+r_2\omega^2
\\
&+P_{t+1}\|p +\theta_p\operatorname{Rot}(\vartheta)v \|^2
\\
&+S_{t+1}(\vartheta+\theta_\vartheta\omega)^2
\\
&+P_{t+1}\operatorname{tr}(\Sigma_p)
+S_{t+1}\sigma_\vartheta^2+c_{t+1}\Bigr\}.
\end{align*}
Define
\begin{align*}
F_t(v ;p ,\vartheta)
&=r_1\|v \|^2
+P_{t+1}\|p +\theta_p\operatorname{Rot}(\vartheta)v \|^2,\\
G_t(\omega;\vartheta)
&=r_2\omega^2
+S_{t+1}(\vartheta+\theta_\vartheta\omega)^2.
\end{align*}
Then
\begin{align*}
V_t(p ,\vartheta)
={}&q_1\|p \|^2+q_2\vartheta^2
+P_{t+1}\operatorname{tr}(\Sigma_p)
\\
&+S_{t+1}\sigma_\vartheta^2+c_{t+1}
\\
&+\min_{v ,\omega}\bigl\{
F_t(v ;p ,\vartheta)
+G_t(\omega;\vartheta)\bigr\}.
\end{align*}
The function $F_t$ depends only on $v $ when $(p ,\vartheta)$ is fixed, and $G_t$ depends only on $\omega$. For all $v ,\omega$,
\begin{align*}
F_t(v )+G_t(\omega)
&\ge\min_z F_t(z)+\min_\eta G_t(\eta).
\end{align*}
Choosing a minimizer of each component attains equality; therefore
\begin{equation*}
{\min_{v ,\omega}(F_t+G_t)
=\min_{v }F_t+\min_\omega G_t.}
\end{equation*}

\subsection{Minimization over the Translational Input}
For fixed $(p,\vartheta)$, write $P=P_{t+1}$. Then
\begin{align*}
F_t(v )
&=r_1\|v \|^2
+P\|p +\theta_p\operatorname{Rot}(\vartheta)v \|^2\\
&=r_1v ^\top v
+P(p +\theta_p\operatorname{Rot}(\vartheta)v )^\top\\
&\hspace{21mm}\cdot(p +\theta_p\operatorname{Rot}(\vartheta)v )\\
&=r_1v ^\top v +P\bigl[p^\top p
+\theta_pp ^\top\operatorname{Rot}(\vartheta)v \\
&\hspace{11mm}+\theta_pv ^\top\operatorname{Rot}(\vartheta)^\top p
+\theta_p^2v ^\top\operatorname{Rot}(\vartheta)^\top\operatorname{Rot}(\vartheta)v \bigr].
\end{align*}
Since
\begin{equation*}
p^\top\operatorname{Rot}(\vartheta)v
=\bigl(p^\top\operatorname{Rot}(\vartheta)v \bigr)^\top
=v^\top\operatorname{Rot}(\vartheta)^\top p .
\end{equation*}
Also, by orthogonality,
\begin{equation*}
v^\top\operatorname{Rot}(\vartheta)^\top\operatorname{Rot}(\vartheta)v
=v^\top I_2v =\|v \|^2.
\end{equation*}
Hence,
\begin{align*}
F_t(v )
={}&r_1\|v \|^2+P\|p \|^2
+2\theta_p Pv ^\top\operatorname{Rot}(\vartheta)^\top p
\\
&+\theta_p^2P\|v \|^2
\\
={}&(r_1+\theta_p^2P)v^\top v
+2\theta_pPv ^\top\operatorname{Rot}(\vartheta)^\top p
\\
&+Pp ^\top p .
\end{align*}

For $v =(v_1,v_2)^\top$, we have
\begin{align*}
\nabla_{v }(v^\top v )
&=\begin{bmatrix}
\partial(v_1^2+v_2^2)/\partial v_1\\
\partial(v_1^2+v_2^2)/\partial v_2
\end{bmatrix}\\
&=\begin{bmatrix}2v_1\\2v_2\end{bmatrix}
=2v .
\end{align*}
For any $b\in\mathbb R^2$ independent of $v $,
\begin{equation*}
\nabla_{v }(v^\top b)=b.
\end{equation*}
Since $\operatorname{Rot}(\vartheta)^\top p $ depends on the \emph{fixed state} but not on the decision $v $,
\begin{align*}
\nabla_{v }F_t(v )
={}&(r_1+\theta_p^2P)\,2v
+2\theta_pP\operatorname{Rot}(\vartheta)^\top p \\
&+\nabla_{v }(Pp ^\top p )\\
={}&2(r_1+\theta_p^2P)v
+2\theta_pP\operatorname{Rot}(\vartheta)^\top p .
\end{align*}
Setting the gradient to zero gives
\begin{align*}
0&=\nabla_{v }F_t(v^{\mathrm{opt}})\\
&=2(r_1+\theta_p^2P)v^{\mathrm{opt}}
+2\theta_pP\operatorname{Rot}(\vartheta)^\top p ,
\end{align*}
which is equivalent to
\begin{align*}
(r_1+\theta_p^2P)v^{\mathrm{opt}}
&=-\theta_pP\operatorname{Rot}(\vartheta)^\top p ,\\
v^{\mathrm{opt}}&=-\frac{\theta_pP}{r_1+\theta_p^2P}
\operatorname{Rot}(\vartheta)^\top p .
\end{align*}
Thus,
\begin{equation*}
{v_t^{\mathrm{opt}}
=-\frac{\theta_pP_{t+1}}{r_1+\theta_p^2P_{t+1}}
\operatorname{Rot}(\vartheta_t)^\top p_t.}
\end{equation*}
In coordinates,
\begin{align*}
\operatorname{Rot}(\vartheta_t)^\top p_t
&=\begin{bmatrix}
\cos\vartheta_t&\sin\vartheta_t\\
-\sin\vartheta_t&\cos\vartheta_t
\end{bmatrix}
\begin{bmatrix}p_{x,t}\\p_{y,t}\end{bmatrix}\\
&=\begin{bmatrix}
\cos\vartheta_t\,p_{x,t}+\sin\vartheta_t\,p_{y,t}\\
-\sin\vartheta_t\,p_{x,t}+\cos\vartheta_t\,p_{y,t}
\end{bmatrix},
\end{align*}
so that
\begin{align*}
v_{1,t}^{\mathrm{opt}}
&=-\frac{\theta_pP_{t+1}}{r_1+\theta_p^2P_{t+1}}
(\cos\vartheta_t\,p_{x,t}+\sin\vartheta_t\,p_{y,t}),\\
v_{2,t}^{\mathrm{opt}}
&=-\frac{\theta_pP_{t+1}}{r_1+\theta_p^2P_{t+1}}
(-\sin\vartheta_t\,p_{x,t}+\cos\vartheta_t\,p_{y,t}).
\end{align*}

The Hessian is
\begin{align*}
\nabla_{v }^2F_t(v )
&=\nabla_{v }\Bigl[
2(r_1+\theta_p^2P)v
+2\theta_pP\operatorname{Rot}(\vartheta)^\top p \Bigr]\\
&=2(r_1+\theta_p^2P)I_2.
\end{align*}
For every nonzero $z\in\mathbb R^2$,
\begin{align*}
z^\top\nabla_{v }^2F_t(v )z
&=2(r_1+\theta_p^2P)z^\top z\\
&=2(r_1+\theta_p^2P)\|z\|^2\\
&\ge2r_1\|z\|^2>0,
\end{align*}
since $r_1>0$ and $P\ge0$. Thus $\nabla_v^2F_t(v)\succ0$ for all $v\in\mathbb R^2$, and $v^{\mathrm{opt}}$ is the unique global minimizer.

Write $D_p=r_1+\theta_p^2P>0$. From the stationary point at $(p,\vartheta)$,
\begin{equation*}
v^{\mathrm{opt}}=-\frac{\theta_pP}{D_p}\operatorname{Rot}(\vartheta)^\top p .
\end{equation*}
By orthogonality,
\begin{align*}
\|v^{\mathrm{opt}}\|^2
&=\frac{\theta_p^2P^2}{D_p^2}
\|\operatorname{Rot}(\vartheta)^\top p \|^2\\
&=\frac{\theta_p^2P^2}{D_p^2}\|p \|^2.
\end{align*}
Also,
\begin{align*}
(v^{\mathrm{opt}})^\top\operatorname{Rot}(\vartheta)^\top p
&=-\frac{\theta_pP}{D_p}
\|\operatorname{Rot}(\vartheta)^\top p \|^2\\
&=-\frac{\theta_pP}{D_p}\|p \|^2.
\end{align*}
Substituting both into the expanded expression for $F_t$ gives
\begin{align*}
F_t(v^{\mathrm{opt}})
={}&D_p\frac{\theta_p^2P^2}{D_p^2}\|p \|^2
+2\theta_pP\left(-\frac{\theta_pP}{D_p}\|p \|^2\right)\\
&+P\|p \|^2\\
={}&\left(\frac{\theta_p^2P^2}{D_p}
-\frac{2\theta_p^2P^2}{D_p}+P\right)\|p \|^2\\
={}&\left(P-\frac{\theta_p^2P^2}{D_p}\right)\|p \|^2\\
={}&\frac{P(D_p-\theta_p^2P)}{D_p}\|p \|^2\\
={}&\frac{r_1P}{r_1+\theta_p^2P}\|p \|^2.
\end{align*}
Therefore,
\begin{equation*}
{\min_{v }F_t(v )
=\frac{r_1P_{t+1}}{r_1+\theta_p^2P_{t+1}}\|p \|^2.}
\end{equation*}

Completing the square gives
\begin{align*}
F_t(v )
={}&D_p\left\|v
+\frac{\theta_pP}{D_p}\operatorname{Rot}(\vartheta)^\top p \right\|^2\\
&+\left(P-\frac{\theta_p^2P^2}{D_p}\right)\|p \|^2\\
={}&D_p\left\|v
+\frac{\theta_pP}{D_p}\operatorname{Rot}(\vartheta)^\top p \right\|^2\\
&+\frac{r_1P}{D_p}\|p \|^2\\
={}&D_p\|v -v^{\mathrm{opt}}\|^2
+\frac{r_1P}{D_p}\|p \|^2.
\end{align*}
Thus, for all $v\in\mathbb R^2$,
\begin{equation*}
F_t(v )-F_t(v^{\mathrm{opt}})
=D_p\|v -v^{\mathrm{opt}}\|^2\ge0,
\end{equation*}
with equality if and only if $v =v^{\mathrm{opt}}$.

\subsection{Minimization over the Angular Input}

For fixed $\vartheta$, write $S=S_{t+1}$. Then
\begin{align*}
G_t(\omega)
&=r_2\omega^2+S(\vartheta+\theta_\vartheta\omega)^2\\
&=r_2\omega^2
+S(\vartheta^2+2\theta_\vartheta\vartheta\omega
+\theta_\vartheta^2\omega^2)\\
&=(r_2+\theta_\vartheta^2S)\omega^2
+2\theta_\vartheta S\vartheta\omega+S\vartheta^2.
\end{align*}

Differentiating gives
\begin{align*}
\frac{\partial G_t}{\partial\omega}
&=2(r_2+\theta_\vartheta^2S)\omega
+2\theta_\vartheta S\vartheta
+\frac{\partial}{\partial\omega}(S\vartheta^2)\\
&=2(r_2+\theta_\vartheta^2S)\omega
+2\theta_\vartheta S\vartheta.
\end{align*}
The stationarity equation becomes
\begin{align*}
0&=\frac{\partial G_t}{\partial\omega}(\omega^{\mathrm{opt}})\\
&=2(r_2+\theta_\vartheta^2S)\omega^{\mathrm{opt}}
+2\theta_\vartheta S\vartheta,\\
(r_2+\theta_\vartheta^2S)\omega^{\mathrm{opt}}
&=-\theta_\vartheta S\vartheta,\\
\omega^{\mathrm{opt}}&=-\frac{\theta_\vartheta S}
{r_2+\theta_\vartheta^2S}\vartheta.
\end{align*}
Hence
\begin{equation*}
{\omega_t^{\mathrm{opt}}
=-\frac{\theta_\vartheta S_{t+1}}
{r_2+\theta_\vartheta^2S_{t+1}}\vartheta_t.}
\end{equation*}

The second derivative is
\begin{align*}
\frac{\partial^2G_t}{\partial\omega^2}
&=\frac{\partial}{\partial\omega}
\Bigl[2(r_2+\theta_\vartheta^2S)\omega
+2\theta_\vartheta S\vartheta\Bigr]\\
&=2(r_2+\theta_\vartheta^2S)\\
&\ge2r_2>0.
\end{align*}
Hence $G_t$ is strictly convex on $\mathbb R$, and $\omega^{\mathrm{opt}}$ is its unique global minimizer.

Write $D_\vartheta=r_2+\theta_\vartheta^2S>0$. Then
\begin{equation*}
\omega^{\mathrm{opt}}=-\frac{\theta_\vartheta S}{D_\vartheta}\vartheta,
\qquad
(\omega^{\mathrm{opt}})^2=\frac{\theta_\vartheta^2S^2}{D_\vartheta^2}\vartheta^2.
\end{equation*}
Substitution into the quadratic gives
\begin{align*}
G_t(\omega^{\mathrm{opt}})
={}&D_\vartheta\frac{\theta_\vartheta^2S^2}{D_\vartheta^2}\vartheta^2
+2\theta_\vartheta S\vartheta
\left(-\frac{\theta_\vartheta S}{D_\vartheta}\vartheta\right)
+S\vartheta^2\\
={}&\left(\frac{\theta_\vartheta^2S^2}{D_\vartheta}
-\frac{2\theta_\vartheta^2S^2}{D_\vartheta}+S\right)\vartheta^2\\
={}&\left(S-\frac{\theta_\vartheta^2S^2}{D_\vartheta}\right)\vartheta^2\\
={}&\frac{S(D_\vartheta-\theta_\vartheta^2S)}{D_\vartheta}\vartheta^2\\
={}&\frac{r_2 S}{r_2+\theta_\vartheta^2S}\vartheta^2.
\end{align*}
Therefore,
\begin{equation*}
{\min_\omega G_t(\omega)
=\frac{r_2 S_{t+1}}
{r_2+\theta_\vartheta^2S_{t+1}}\vartheta^2.}
\end{equation*}

Completing the square yields
\begin{align*}
G_t(\omega)
={}&D_\vartheta\left(\omega
+\frac{\theta_\vartheta S}{D_\vartheta}\vartheta\right)^2
+\left(S-\frac{\theta_\vartheta^2S^2}{D_\vartheta}\right)\vartheta^2\\
={}&D_\vartheta(\omega-\omega^{\mathrm{opt}})^2
+\frac{r_2 S}{D_\vartheta}\vartheta^2.
\end{align*}
In particular,
\begin{equation*}
G_t(\omega)-G_t(\omega^{\mathrm{opt}})
=D_\vartheta(\omega-\omega^{\mathrm{opt}})^2\ge0,
\end{equation*}
with equality only at $\omega=\omega^{\mathrm{opt}}$.

With $P=P_{t+1}$ and $S=S_{t+1}$, define
\begin{equation*}
H_t(v ,\omega)=F_t(v )+G_t(\omega).
\end{equation*}
For $u=(v_1,v_2,\omega)^\top$,
\begin{equation*}
\nabla_uH_t=
\begin{bmatrix}
2(r_1+\theta_p^2P)v +2\theta_pP\operatorname{Rot}(\vartheta)^\top p \\
2(r_2+\theta_\vartheta^2S)\omega+2\theta_\vartheta S\vartheta
\end{bmatrix}.
\end{equation*}
There are no cross derivatives between $v $ and $\omega$:
\begin{equation*}
\nabla_{v }\frac{\partial G_t}{\partial\omega}=0,
\qquad
\frac{\partial}{\partial\omega}\nabla_{v }F_t=0.
\end{equation*}
Therefore the full Hessian is
\begin{equation*}
\nabla_u^2H_t
=2\begin{bmatrix}
(r_1+\theta_p^2P)I_2&0\\
0&r_2+\theta_\vartheta^2S
\end{bmatrix}.
\end{equation*}
For every nonzero $(z,\eta)\in\mathbb R^2\times\mathbb R$,
\begin{align*}
&\begin{bmatrix}z^\top&\eta\end{bmatrix}
\nabla_u^2H_t\begin{bmatrix}z\\\eta\end{bmatrix}\\
&=2(r_1+\theta_p^2P)\|z\|^2
+2(r_2+\theta_\vartheta^2S)\eta^2\\
&\ge2\min\{r_1,r_2\}\bigl(\|z\|^2+\eta^2\bigr)>0.
\end{align*}
Thus $H_t$ is strongly convex in $u$ for each fixed state. By the preceding square completions,
\begin{align*}
&H_t(v ,\omega)-H_t(v^{\mathrm{opt}},\omega^{\mathrm{opt}})
\\
&=(r_1+\theta_p^2P)\|v -v^{\mathrm{opt}}\|^2
+(r_2+\theta_\vartheta^2S)(\omega-\omega^{\mathrm{opt}})^2
\ge0.
\end{align*}

\subsection{Backward Recursion}
Substitute the minimized values $\min_v F_t(v)$ and $\min_\omega G_t(\omega)$ into the separated Bellman equation:
\begin{align*}
V_t(p ,\vartheta)
={}&q_1\|p \|^2+q_2\vartheta^2
+P_{t+1}\operatorname{tr}(\Sigma_p)\\
&+S_{t+1}\sigma_\vartheta^2+c_{t+1}\\
&+\frac{r_1P_{t+1}}{r_1+\theta_p^2P_{t+1}}\|p \|^2\\
&+\frac{r_2 S_{t+1}}
{r_2+\theta_\vartheta^2S_{t+1}}\vartheta^2\\
={}&\left(q_1+\frac{r_1P_{t+1}}
{r_1+\theta_p^2P_{t+1}}\right)\|p \|^2\\
&+\left(q_2+\frac{r_2 S_{t+1}}
{r_2+\theta_\vartheta^2S_{t+1}}\right)\vartheta^2\\
&+c_{t+1}+P_{t+1}\operatorname{tr}(\Sigma_p)
+S_{t+1}\sigma_\vartheta^2.
\end{align*}
Hence,
\begin{equation*}
{V_t(p ,\vartheta)
=P_t\|p \|^2+S_t\vartheta^2+c_t,}
\end{equation*}
where, for $t=T-1,\ldots,0$,
\begin{align*}
{P_t
=q_1+\frac{r_1P_{t+1}}
{r_1+\theta_p^2P_{t+1}}},
\\
{S_t
=q_2+\frac{r_2 S_{t+1}}
{r_2+\theta_\vartheta^2S_{t+1}}},
\\
{c_t=c_{t+1}
+P_{t+1}\operatorname{tr}(\Sigma_p)
+S_{t+1}\sigma_\vartheta^2.}
\end{align*}

The terminal values are $P_T=q_1$, $S_T=q_2$, and $c_T=0$.

At time $T$, $P_T=q_1>0$ and $S_T=q_2>0$. If $P_{t+1}>0$ and $S_{t+1}>0$, then
\begin{align*}
r_1+\theta_p^2P_{t+1}&\ge r_1>0,\\
r_2+\theta_\vartheta^2S_{t+1}&\ge r_2>0.
\end{align*}
Moreover,
\begin{align*}
P_t&=q_1+\frac{r_1P_{t+1}}
{r_1+\theta_p^2P_{t+1}}\ge q_1>0,\\
S_t&=q_2+\frac{r_2 S_{t+1}}
{r_2+\theta_\vartheta^2S_{t+1}}\ge q_2>0.
\end{align*}
Since all noise variances are nonnegative, $c_t\ge c_{t+1}\ge0$. Backward induction proves that the quadratic value-function form and the positivity properties hold for every $t=0,\ldots,T$.

Since $c_T=0$, the recursion for the additive constant gives
\begin{equation*}
 c_t=\sum_{k=t}^{T-1}
 \bigl(P_{k+1}\operatorname{tr}(\Sigma_p)+S_{k+1}\sigma_\vartheta^2\bigr).
\end{equation*}

\subsection{Global Optimality}
The optimal state-feedback controller is
\begin{equation*}
{\begin{aligned}
v_t^{\mathrm{opt}}&=-\frac{\theta_pP_{t+1}}
{r_1+\theta_p^2P_{t+1}}
\operatorname{Rot}(\vartheta_t)^\top p_t,\\
\omega_t^{\mathrm{opt}}&=-\frac{\theta_\vartheta S_{t+1}}
{r_2+\theta_\vartheta^2S_{t+1}}\vartheta_t.
\end{aligned}}
\end{equation*}
Define
\begin{align*}
 k_{p,t}^{\mathrm{opt}}(\theta)&=\frac{\theta_p P_{t+1}}{r_1+\theta_p^2P_{t+1}},\\
 k_{\vartheta,t}^{\mathrm{opt}}(\theta)&=\frac{\theta_\vartheta S_{t+1}}{r_2+\theta_\vartheta^2S_{t+1}}.
\end{align*}
Thus the controller in~\eqref{eq:robot_optimal} is
\begin{equation*}
 {\pi_t^{\mathrm{opt}}(x;\theta)=
 \begin{bmatrix}
 -k_{p,t}^{\mathrm{opt}}(\theta)\operatorname{Rot}(\vartheta)^\top p\\
 -k_{\vartheta,t}^{\mathrm{opt}}(\theta)\vartheta
 \end{bmatrix}.}
\end{equation*}

Let $\ell_t(x_t,u_t)=x_t^\top Qx_t+u_t^\top Ru_t$. For any causal controller with $\mathcal F_t$-measurable actions and finite expected cost, let $(v_t^{\mathrm{opt}},\omega_t^{\mathrm{opt}})$ be the optimal feedback evaluated at its current state $x_t$. By completing the squares,
\begin{align*}
 &\ell_t(x_t,u_t)+\mathbb E[V_{t+1}(p_{t+1},\vartheta_{t+1})\mid\mathcal F_t]\\
 &\quad-V_t(p_t,\vartheta_t)\\
 &\quad=(r_1+\theta_p^2P_{t+1})\|v_t-v_t^{\mathrm{opt}}\|^2\\
 &\qquad+(r_2+\theta_\vartheta^2S_{t+1})(\omega_t-\omega_t^{\mathrm{opt}})^2.
\end{align*}
Taking expectations conditional on $\theta$ and summing over $t$, the tower property gives
\begin{align*}
 &\sum_{t=0}^{T-1}\mathbb E\Big[
 \mathbb E[V_{t+1}(p_{t+1},\vartheta_{t+1})\mid\mathcal F_t]\\
 &\hspace{23mm}-V_t(p_t,\vartheta_t)\mid\theta\Big]\\
 &\quad=\sum_{t=0}^{T-1}\Big(
 \mathbb E[V_{t+1}(p_{t+1},\vartheta_{t+1})\mid\theta]\\
 &\hspace{23mm}-\mathbb E[V_t(p_t,\vartheta_t)\mid\theta]\Big)\\
 &\quad=\mathbb E[V_T(p_T,\vartheta_T)\mid\theta]
 -\mathbb E[V_0(p_0,\vartheta_0)\mid\theta].
\end{align*}
Consequently,
\begin{align*}
 &\mathbb E \Big[\sum_{t=0}^{T-1}\ell_t(x_t,u_t)+V_T(p_T,\vartheta_T)\\
 &\hspace{20mm}-V_0(p_0,\vartheta_0)\mid\theta\Big]\\
 &\quad=\sum_{t=0}^{T-1}\mathbb E \Big[(r_1+\theta_p^2P_{t+1})\|v_t-v_t^{\mathrm{opt}}\|^2\\
 &\qquad+(r_2+\theta_\vartheta^2S_{t+1})(\omega_t-\omega_t^{\mathrm{opt}})^2\mid\theta\Big].
\end{align*}
Since $V_T(p_T,\vartheta_T)=x_T^\top Qx_T$, this is
\begin{align*}
 &J(\pi;\theta)-\mathbb E[V_0(p_0,\vartheta_0)\mid\theta]\\
 &\quad=\sum_{t=0}^{T-1}\mathbb E \Big[(r_1+\theta_p^2P_{t+1})\|v_t-v_t^{\mathrm{opt}}\|^2\\
 &\qquad+(r_2+\theta_\vartheta^2S_{t+1})(\omega_t-\omega_t^{\mathrm{opt}})^2\mid\theta\Big]\ge0.
\end{align*}
Under $\pi^{\mathrm{opt}}$, the feedback gains are finite and
$\operatorname{Rot}(\vartheta_t)$ is orthogonal. Hence the controls have finite
second moments whenever the state does. The dynamics and the finite noise
second moments then imply, by induction from $x_0$, that all state and control
second moments are finite. Thus $J(\pi^{\mathrm{opt}};\theta)<\infty$.
For $\pi=\pi^{\mathrm{opt}}$ the right-hand side is zero. Hence
\begin{equation*}
 J(\pi^{\mathrm{opt}};\theta)=\mathbb E[V_0(p_0,\vartheta_0)\mid\theta]\le J(\pi;\theta).
\end{equation*}
Since both coefficients are strictly positive, equality holds only if the actions agree with the optimal feedback almost surely along the trajectory. Since the optimal cost is finite, the conclusion also holds for policies of infinite cost. This proves the global optimality of the feedback controller in~\eqref{eq:robot_optimal}.\hfill$\square$

\subsection{Verification of the Assumptions}
For the robot dynamics and the nominal-gain controller library in
Section~\ref{sec:experiments}, we verify the conditions of
Theorem~\ref{thm:nonlinear_finite} as follows.

\noindent\textit{Assumption~\ref{ass:system}.}
For $\theta_p,\theta_\vartheta\in[0.55,1.55]$, the orthogonality of
$\operatorname{Rot}(\vartheta)$ gives
\begin{align*}
 \|f(x,u,w;\theta)\|
 &=\left\|x+\begin{bmatrix}
 \theta_p\operatorname{Rot}(\vartheta)v\\
 \theta_\vartheta\omega
 \end{bmatrix}+w\right\|\\
 &\le\|x\|+\left(\theta_p^2\|v\|^2
 +\theta_\vartheta^2\omega^2\right)^{1/2}+\|w\|\\
 &\le\|x\|+1.55\left(\|v\|^2+\omega^2\right)^{1/2}+\|w\|\\
 &=\|x\|+1.55\|u\|+\|w\|.
\end{align*}
Thus, $a_0=0$, $a_1=1$, $a_2=1.55$, $a_3=1$, and
$p_x=p_u=p_w=1$, uniformly over the parameter support.

\noindent\textit{Assumption~\ref{ass:controller}.}
For the nominal-gain controller library, write
$k_{p,t}^{(j)}=\alpha_p^{(j)}k_{p,t}^{\mathrm{nom}}$ and
$k_{\vartheta,t}^{(j)}=\alpha_\vartheta^{(j)}k_{\vartheta,t}^{\mathrm{nom}}$.
Then
\begin{align*}
 \|\pi_t^{(j)}(x)\|^2
 &=|k_{p,t}^{(j)}|^2\|\operatorname{Rot}(\vartheta)^\top p\|^2
 +|k_{\vartheta,t}^{(j)}|^2\vartheta^2\\
 &=|k_{p,t}^{(j)}|^2\|p\|^2
 +|k_{\vartheta,t}^{(j)}|^2\vartheta^2\\
 &\le a_5(T)^2(\|p\|^2+\vartheta^2)\\
 &=a_5(T)^2\|x\|^2,
\end{align*}
where, as in Section~\ref{sec:experiments},
\begin{equation*}
 a_5(T)=\max_{\substack{j=1,\ldots,L\\t=0,\ldots,T-1}}
 \left\{|k_{p,t}^{(j)}|,|k_{\vartheta,t}^{(j)}|\right\}<\infty.
\end{equation*}
Taking square roots gives $\|\pi_t^{(j)}(x)\|\le a_5(T)\|x\|$.
Hence $a_4=0$, $a_5=a_5(T)$, and $p_\pi=1$.

\noindent\textit{Assumption~\ref{ass:noise}.}
Since $w_t\sim\mathcal N(0,0.09I_3)$, each coordinate satisfies
$w_{t,j}\sim\mathcal N(0,0.09)$ and
\begin{equation*}
 \mathbb E\exp(\eta w_{t,j})
 =\exp\left(\frac{0.09\eta^2}{2}\right)
 =\exp\left(\frac{\eta^2\sigma_w^2}{2}\right),
 \quad \eta\in\mathbb R,
\end{equation*}
with $\sigma_w=0.3$.

Finally, since all growth exponents equal one, $d=b=1$ and $h=4$.
For the deterministic initial state $x_0=(2,-1.5,1.15)^\top$,
\begin{align*}
 \mathbb E(1+\|x_0\|)^h
 &=\left(1+\sqrt{2^2+(-1.5)^2+1.15^2}\right)^4\\
 &=M_0<\infty.
\end{align*}
Therefore the initial-state moment condition also holds.

\bibliographystyle{IEEEtran}
\bibliography{references}

\end{document}